\documentclass[3p,twocolumn]{elsarticle}

\usepackage{lineno,hyperref}
\modulolinenumbers[5]
\usepackage{graphicx}
\usepackage{float}
\usepackage{multirow}
\usepackage{longtable}
\usepackage{ulem}
\usepackage{natbib}
\usepackage{xcolor}
\usepackage{amssymb}
\usepackage{amsmath}
\usepackage{adjustbox}
\setcitestyle{square, comma, numbers,sort&compress, super}
\newcommand{\cm}{\ensuremath{\textrm{cm}^{-1}}}
\newcommand{\K}{K$^{-1}$}

\newcommand{\km}{\ensuremath{\textrm{km}\cdot\textrm{mol}^{-1} cm}}

\journal{Journal of Molecular Spectroscopy}

\begin{document}

\begin{frontmatter}

\title{Anharmonic Infrared Spectra of Thermally Excited Pyrene (C$_{16}$H$_{10}$): A Combined View of DFT-Based GVPT2 with AnharmonicCaOs, and Approximate DFT Molecular dynamics with DemonNano}


\author[mymainaddress]{Shubhadip Chakraborty}

\author[mymainaddress,mysecondaryaddress]{Giacomo Mulas\corref{mycorrespondingauthor}}

\author[mysecondaryaddress1]{Mathias Rapacioli}

\author[mymainaddress]{Christine Joblin}

\cortext[mycorrespondingauthor]{Corresponding author}

\address[mymainaddress]{Institut de Recherche en Astrophysique et Plan\'etologie, Universit\'e de Toulouse (UPS), CNRS, CNES, 9 Av. du Colonel Roche, 31028 Toulouse Cedex 4, France}
\address[mysecondaryaddress]{Istituto Nazionale di Astrofisica (INAF), Osservatorio Astronomico di Cagliari, 09047 Selargius (CA), Italy}
\address[mysecondaryaddress1]{Laboratoire de Chimie et Physique Quantiques (LCPQ/IRSAMC), Universit\'e de Toulouse (UPS),CNRS, 118 Route de Narbonne, 31062 Toulouse, France}

\begin{abstract}
The study of the Aromatic Infrared Bands (AIBs) in astronomical environments has opened interesting spectroscopic questions on the effect of anharmonicity on the infrared (IR) spectrum of hot polycyclic aromatic hydrocarbons (PAHs) and related species in isolated conditions. The forthcoming James Webb Space Telescope will unveil unprecedented spatial and spectral details in the AIB spectrum; significant advancement is thus necessary \textit{now} to model the infrared emission of PAHs, their presumed carriers, with enough detail to exploit the information content of the AIBs. This requires including anharmonicity in such models, and to do so systematically for all species included, requiring a difficult compromise between accuracy and efficiency.\\ 
We performed a benchmark study to compare the performances of two methods in calculating anharmonic spectra, comparing them to available experimental data. One is a full quantum method, AnharmoniCaOs, relying on an {\it ab initio} potential, and the other relies on Molecular Dynamics simulations using a Density Functional based Tight Binding potential.
The first one is found to be very accurate and detailed, but it becomes computationally very expensive for increasing temperature; the second is faster and can be used for arbitrarily high temperatures, but is less accurate. Still, its results can be used to model the evolution with temperature of isolated bands. \\
We propose a new recipe to model anharmonic AIB emission using minimal assumptions on the general behaviour of band positions and widths with temperature, which can be defined by a small number of empirical parameters. Modelling accuracy will depend critically on these empirical parameters, allowing for an incremental improvement in model results, as better estimates become gradually available.

\end{abstract}

\begin{keyword}
\texttt{Anharmonic Infrared Spectroscopy \sep Polycyclic Aromatic Hydrocarbons \sep Quantum Chemistry \sep Molecular dynamics \sep Astrochemistry}
\end{keyword}

\end{frontmatter}


\section{Introduction}
Aromatic Infrared Bands (AIBs) are a set of bright and ubiquitous emission bands, observed in regions illuminated by stellar ultraviolet (UV) photons, from our galaxy all the way out to cosmological distances. Their main components fall at $\sim$3.3, 6.2, 7.7, 8.6, 11.3, 12.7, and 16.4\,$\mu$m. PAH species constitute the most plausible candidate carriers since they have IR bands which correspond at least to first order to the AIBs and they can reach high temperatures (1000\,K or more) upon the absorption of a single UV photon due to their limited number of degrees of freedom. Because of the extreme isolation in space, the hot molecules can then relax by slow infrared emission \citep{Leger1989}. \\
The coming James Webb Space Telescope (JWST) will provide a wealth of new data on the AIBs and their spatial variation, gathering a mine of information about the chemical identity of the emitting PAHs and their chemical evolution in various environments. However access to this mine will be conditioned by our capacity to produce relevant synthetic spectra with a model that can describe the photophysics of a given PAH molecule in a given UV-visible astrophysical radiation field \citep{Mulas2006a}. For a comparison with astronomical spectra, such a model needs to take into account anharmonic effects while simulating the IR emission spectra. However the effect of anharmonicity is usually included in a simplistic way using an average band shift (typically $\sim$10\,cm$^{-1}$) and broadening (typically $\sim$10-30\,cm$^{-1}$) (cf. for instance the tool available in the AmesPAHdbIDLSuite in order to calculate an emission spectrum from a theoretical IR spectrum as available in the NASA Ames PAH IR Spectroscopic Database \cite{bauschlicher2018, Mattioda2020}).
So far, only a couple of models have tried to go beyond these approximations while simulating the cooling of hot PAHs to match astronomical observations \citep{Cook98_model,verstraete2001,Pech2002,Mulas2006b}. The first three models were based on empirical parameters that quantify the linear slopes derived from the evolution of band positions with temperature in the  spectra of thermally excited PAHs \citep{Joblin1995}. The last model by \citet{Mulas2006b} was based only on theory and focused on the case of the small molecules, namely naphthalene (C$_{10}$H$_8$) and anthracene (C$_{14}$H$_{10}$), for which the matrix of anharmonic parameters of the Dunham expansion expressing vibrational energy could be calculated, and included in the Monte Carlo simulation. With the coming JWST, it appears very timely to consider these effects in a more systematic way. \\
In the laboratory, several techniques were used  to quantify anharmonicity. The evolution of the IR absorption spectra of neutral gas-phase PAHs at high temperatures was studied using thermal excitation in gas cells \citep{Kurtz1992, Joblin1995}. In the case of PAH ions, other techniques have to be used and \citet{Oomens2003} discussed the potential of IR multiple photon dissociation (IRMPD) action spectroscopy of trapped ions. Although IRMPD spectra contain non-linearities and cannot be considered as representative of IR absorption or emission spectra, they carry information on the impact of anharmonicity on band positions and widths. The technique can be conveniently used for large PAH ions of astrophysical interest \citep{Zhen2018} to provide at a first order IR spectra of PAHs containing an internal energy comparable to that of astro-PAHs absorbing UV photons. Experiments recording the IR emission spectra of small UV-excited PAHs were performed in the early 1990s \citep{Cherchneff1989,Shan1991, Brenner1992}. This work performed on the 3.3\,$\mu$m band of small PAHs (up to pyrene in the case of \citet{Shan1991}) was extended to the full mid-IR range thanks to the IR photon-counting experiment developed in the Saykally's group \citep{Schlemmer94,Cook98}. These difficult experiments demonstrated the validity of the PAH emission model in astrophysical environments. They illustrated that anharmonic effects redshift band positions and lead to broadened and asymmetric band profiles. They constitute a very valuable dataset, provided one can precisely quantify the excitation and de-excitation conditions (role of collisions?) in the experiments as discussed in \citet{Cook98}. \\
So far, the experimental measurements performed on thermally excited gas-phase PAHs have provided the easiest way to quantify anharmonicity for astrophysical models \citep{Cook98_model, Pech2002}. This is due to the fact that the evolution  with temperature of the band positions and widths was found to be linear in the studied range ($\sim$600-900\,K) allowing us to derive empirical anharmonic parameters \citep{Joblin1995}. However these measurements are limited both in the number of species that could be studied but also in the data that they can provide. A recent detailed study on pyrene in pellets covered the full 14--723\,K range \citep{chakraborty2019} and showed that the linear trend observed at high temperatures is not effective all the way down to low temperatures. Another disadvantage of high-temperature gas-phase measurements is that the bandwidths include a contribution from rotational broadening that was removed in the model by \citet{Pech2002}, considering that in astrophysical environments the vibrational and rotational temperatures are not thermalized, with a typical value of the rotation temperature of 100-150\,K \citep{Rouan1992, Mulas1998,Malloci2005}.\\
On the theoretical/modelling side, several models were used/developed to quantify anharmonic effects in the IR spectra of PAHs. They fall in two main categories: one of the possible approaches is to use Molecular Dynamics, based on some kind of \textit{ab initio} engine (e.g. DFT or DFTB) for on the fly evaluation of the potential energy and electric dipole (and/or higher electric/magnetic multipole) hypersurfaces. This method treats nuclear motion classically, thereby losing information on the discrete structure of individual quantised vibrational states. However, it does not make any \textit{a priori} assumption on selection rules, merely following the time evolution of physical quantities and obtaining normal frequencies, transition energies and intensities from the numerical autocorrelation function of the appropriate function. It therefore naturally obtains also combination/overtone bands, provided one accumulates enough data through many, long enough simulations. Achieving ergodicity at relatively low energies can be challenging though, making this method better suited for T$\gtrsim$500~K. This method was used with some success in the context of the investigation of interstellar PAHs spectral properties \cite{parneix,Rapacioli2015,Joalland2010,Simon2011, SimonPCCP2012,WaterMat_JPCA2015}. \\
The second approach is that of a full quantum calculation, the most successful of which, for PAHs, was the Van Vleck formalism of 2$^\mathrm{nd}$ order perturbation theory, with explicit variational treatment of resonances, based on a 4$^\mathrm{th}$ order Taylor expansion of the PES at a stable geometry and a 2$^\mathrm{nd}$ order Taylor expansion of the electric dipole (and/or higher electric/magnetic multipole) moment \citep{piccardo2015}. The Taylor expansions are obtained via \textit{ab initio} calculations, e.~g. with DFT. This representation of the PES and multipole moments makes this method best suited for fairly rigid molecules and/or not too high vibrational excitation, so that these approximations remain good for the vast majority of the accessible phase space. This method is thus very well suited for PAHs at not too high temperatures. The main disadvantage is that adequately sampling phase space with this method becomes computationally very expensive at moderately high temperatures, due to the huge number of resonating states, resulting in huge effective Hamiltonians to be built and diagonalised, and in a correspondingly huge number of dipole matrix elements between their eigenvectors to be computed. This method was successfully applied to naphthalene \citep{pirali2009,basire2009} and, more recently to slightly larger PAHs \citep[see e.~g.][]{lemmens2019, mackie2016, Calvo:2011uq}.

We developed a quantum chemistry code, AnharmoniCaOs, to study the detailed anharmonic spectra of PAHs with a fully quantum approach \citep{Mulas2018}. A first milestone for this code has been to model the effects of anharmonicity at 0\,K, including the effect on the position of the fundamental bands but also the contribution of the overtone, combination, and difference bands to the spectra. The results of the code were compared to available experimental absorption data on pyrene (C$_{16}$H$_{10}$) and coronene (C$_{24}$H$_{12}$). While new experimental absorption data in the gas-phase and at low temperatures are becoming available \citep{Maltseva15, Maltseva16, lemmens2019}, this code can support the interpretation of such experimental data and, once calibrated and validated against them, extend calculations to spectra which are more difficult to measure in the laboratory, e.g. vibrational spectra of ions and/or radicals, in absorption as well as in emission. In this article the focus is to evaluate the ability of this code in modelling absorption spectra at higher temperatures. 
We expect that at temperatures well above $\sim$500\,K, the system will spend an increasingly larger part of its time in regions of the potential energy surface far from the optimised geometry, which cannot be anymore described by a quartic scheme. In order to access anharmonicity factors in such cases, we test the alternative approach given by molecular dynamics, in which the IR spectrum is extracted from a Fourier transform of the dipole autocorrelation function. As millions of energy and forces calculations are mandatory for such scheme, the potential energy will be computed from a parameterized method, namely the Density Functional based Tight Binding (DFTB), an approximated DFT scheme with a much lower computational cost. This approach has already been shown to provide reasonable qualitative and quantitative trends of anharmonicity factors for some specific bands of PAH based systems  \citep{Joalland2010,Simon2011}.\\
This article compiles the results which have been obtained on the evolution of the IR absorption band positions and widths for pyrene as derived from two different modelling approaches, the ab initio AnharmoniCaOs code (henceforth DFT-AC) for the low temperature range (up to $\sim$500\,K) and the DFTB-based MD simulations (henceforth DFTB-MD) for the higher temperatures. The obtained results are compared with available experimental data. This is the first study of the kind dedicated to anharmonic effects in the IR spectra of PAHs in which the performances of these two modelling methods are assessed in relation with known experimental data. We then assess which parameters can be derived from these combined techniques for simplified description of anharmonicity in astrophysical models, providing a schematic recipe. Finally we discuss some perspectives in view of the forthcoming James Webb Space Telescope.

\section{Methodology}\label{S:2}

\subsection{AnharmoniCaOs}
\label{SS:2}
To obtain anharmonic vibrational spectra of pyrene at non-zero temperature, we made use of the AnharmoniCaOs code, developed by some of the authors as described in \citet{Mulas2018}. In brief, AnharmoniCaOs takes in input the harmonic frequencies of all normal modes, the third and fourth derivatives of the potential energy with respect to normal coordinates, as well as the first and second derivatives of the electric dipole moment, and obtains anharmonic states and electric dipole-permitted transitions.  The Taylor expansions for pyrene were taken from \citet{Mulas2018}. AnharmoniCaOs then proceeds with a mixed variational-perturbative approach using the Van Vleck formalism (GVPT2): resonant terms are considered explicitly, building effective Hamiltonian matrices and numerically solving them to obtain anharmonic eigenstates, while non-resonant terms are dealt with by perturbation theory. Anharmonic eigenstates are represented as linear combinations of harmonic states. Matrix elements of the electric dipole operator are then obtained between the anharmonic eigenstates obtained from the diagonalization of the effective Hamiltonians. \\
The set of harmonic states included in each calculation is built around a given harmonic ``start state'', which results in the formation of truncated polyads of states connected by resonances. For more details on the algorithms used to build and truncate polyads, see \citet{Mulas2018}. As a result of the way polyads are built and truncated, one finds the anharmonic states with a significant component along the initial ``starting state''. States with the largest projection along the ``starting state'' are numerically most accurate, while accuracy degrades as anharmonic states get farther and farther ``away'' (in terms of projection) from the ``starting state''. For each of these calculations, therefore, only the anharmonic states closest to the ``starting state'' are considered in building the spectra, with a weight proportional to the squared modulus of their projection along the ``starting state''. \\
If one were to repeat the calculation for all possible ``starting states'', one would obtain all anharmonic states and transitions. However, the density of harmonic states, for a polyatomic molecule, is an extremely steep function of vibrational energy, making complete sampling of harmonic states unfeasible above a relatively low energy. We therefore enumerated harmonic states, in order of harmonic energy, till we reached a predetermined maximum number. From that harmonic energy on ($\sim$1200~cm$^{-1}$),  we resorted to statistical sampling. We used the Wang-Landau method \citep{wang2001a,wang2001b} to perform a random walk on harmonic states, obtaining a nearly uniform sampling in \textit{harmonic} energy space of the ``starting states''. This does not result in a completely uniform sampling in the \textit{anharmonic} energy space for the states obtained by AnharmoniCaOs; still, this produces a very slowly varying sampling density in anharmonic energy space, ensuring that all energy ranges randomly explored are evenly sampled. This sampling method was successfully used, e.~g., by \citet{basire2009} to estimate the anharmonic density of states of naphthalene.\\
The collection of these AnharmoniCaOs runs produces anharmonic vibrational spectra as a function of vibrational energy: the vibrational energy is discretised in a number of bins, each covering an interval of 5~cm$^{-1}$; individual spectra consisting of transitions from an anharmonic state in that interval are averaged, to estimate the spectrum of molecules in thermal equilibrium at the corresponding vibrational energy, in the microcanonical ensemble; these spectra can then undergo a numerical Laplace transform to finally obtain anharmonic vibrational spectra as a function of temperature in the canonical ensemble, directly comparable with laboratory experiments.

\subsection{DFTB simulations}
\label{SS:3}
We have also computed the anharmonic spectra for pyrene from a dynamical exploration of the PES at various temperatures. Such dynamics requires the computation of millions of single point energies and gradients and could hardly be performed at the {\it ab initio} level. A good compromise relies on the use of Density Functional based Tight Binding (DFTB)\cite{dftb2,dftb1}, which presents a much cheaper computational cost than DFT  while preserving a quantum description of the electronic system and allowing for the calculation of the molecular dipole $\vec{\mu}(t)$ on the fly.
In practice, we used the DFTB scheme in its second order formulation \cite{scc-dftb} implemented in deMonNano  \cite{demon}  with the MAT set of parameters \cite{matsci} combined with a correction for atomic charges \cite{DFTB_CM3} that already showed its efficiency in the context of IR spectra calculations \cite{Joalland2010,Rapacioli2015,SimonPCCP2012,WaterMat_JPCA2015}. The IR spectra were obtained from the Fourier transform of the auto-correlation dipole moment using the formulation presented by \cite{gai03} : 
\begin{equation}
\label{Eq:specdyn}
\alpha(\omega)\ \ \propto\ \ \omega^2\ \int_{-\infty}^{+\infty}\textrm{d}t\ \langle\vec{\mu}(0) \cdot \vec{\mu}(t)\rangle\ e\,^{i\omega t}
\end{equation}
 where $<>$ indicates a statistical average to remove dependency on the initial conditions. 

As the equation is intended to be used to compute the spectum at a given temperature, the simulations should be done in the canonical ensemble, which in practice means using a thermostat. The latter might however introduce errors as the process of energy exchange between the system and the thermostat involves frequencies which might pollute the computed spectrum. To reduce such artifacts, we first perform, for each of the 6 investigated temperatures between 600 and 1600\,K, a canonical dynamics of 50~ps using a Nose Hoover chain of thermostat \cite{Hoover1985,Nose1984,Tuckermana1992} (5 thermostats with energy exchange frequency of 800 cm$^{-1}$ ). About 50 snapshots are selected along this dynamics and the corresponding geometries and velocities are used to generate the starting conditions for microcanonical simulations of 5ps each with timestep of 0.1 fs. The IR spectrum is computed and averaged from these last simulations without thermostats, the initial conditions sampling however the canonical distribution at a given temperature.

\subsection{Spectra analysis}\label{sec:analysis}

One of the aims of this work was to obtain, from more or less complete theoretical calculations, the evolution of band positions and bandwidths as a function of temperature, for use in simplified modelling of AIB emission in astronomical environments, comparing results from different theoretical methods and experimental ones. This poses a fundamental problem: how to define in an unambiguous way band positions and widths, reducing to a minimum their dependence on subjective choices of the scientist deriving them from the data. This is particularly difficult for bands that result from the blending of several components, which sometimes can be resolved only at low temperatures. We therefore decided to define the centroid $\bar{\nu}_i$ of the band $i$ as the mathematical average of its position, weighted by band intensity $I(\nu)$, over a fiducial interval $[\nu_i^\mathrm{min}, \nu_i^\mathrm{max}]$ defined by the scientist for that band, i.~e.
\begin{equation}
    \bar{\nu}_i = \frac{ \int_{\nu_i^\mathrm{min}}^{\nu_i^\mathrm{max}} \mathrm{d}\nu \, \nu \, I(\nu)}{\int_{\nu_i^\mathrm{min}}^{\nu_i^\mathrm{max}} \mathrm{d}\nu \, I(\nu)} = \frac{1}{I_i} \int_{\nu_i^\mathrm{min}}^{\nu_i^\mathrm{max}} \mathrm{d}\nu \, \nu \, I(\nu), 
\label{eq:bandcentroid}
\end{equation}
where $I_i$ is the integrated intensity of band $i$.
Similarly, we define the bandwidth $\sigma_i$ as the square root of
\begin{equation}
    \sigma_i^2 = \frac{1}{I_i} \int_{\nu_i^\mathrm{min}}^{\nu_i^\mathrm{max}} \mathrm{d}\nu \, \left( \nu - \bar{\nu}_i \right)^2 \, I(\nu).
\label{eq:bandwidth}
\end{equation}
These definitions are reminiscent of the definition of the first two central momenta of a statistical distribution function.
They are identically applicable to DFT-AC spectra, DFTB-MD, and experimental ones, making it possible to meaningfully compare positions and widths, thus defined, obtained from all three sets of data. It is also evident from the equations defining them that they are independent of the units the spectra are expressed in, and of their normalisation. In particular, they apply both to calibrated spectra in absolute units (e.~g. km mol$^{-1}$ cm) and to spectra in arbitrary units (e.~g. normalised to unity peak intensity).
The integration intervals $[\nu_i^\mathrm{min}, \nu_i^\mathrm{max}]$ for each band were defined by visual inspection of the spectra, and are the only remaining subjective choice. The ones we adopted are given in Tables~S1 and S2 in the supplementary material.

\section{Results}\label{sec:results}

\subsection{AnharmoniCaOs results}
\begin{figure*}
\centering
\includegraphics[width=13cm]{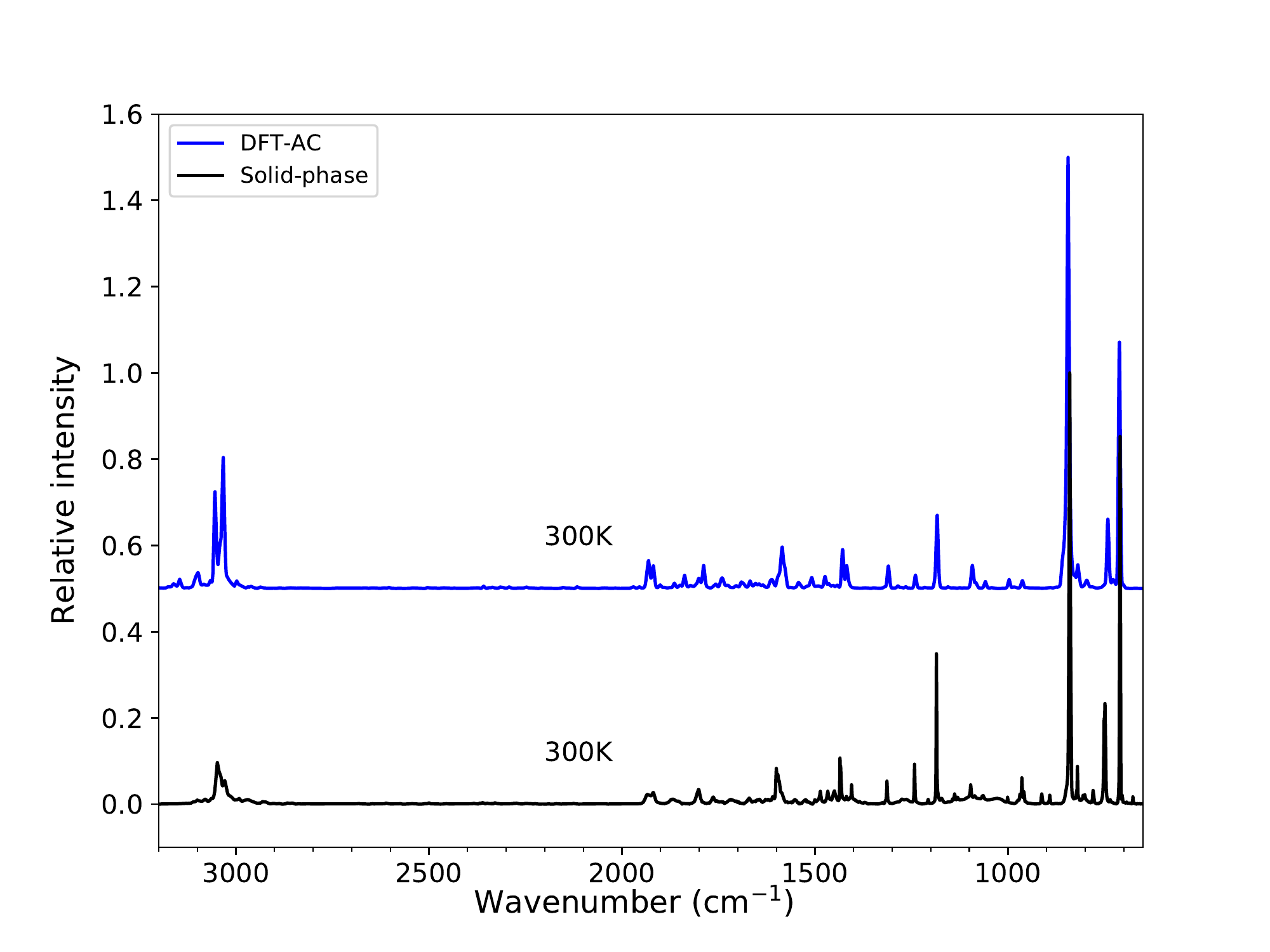} \\ \includegraphics[width=13cm]{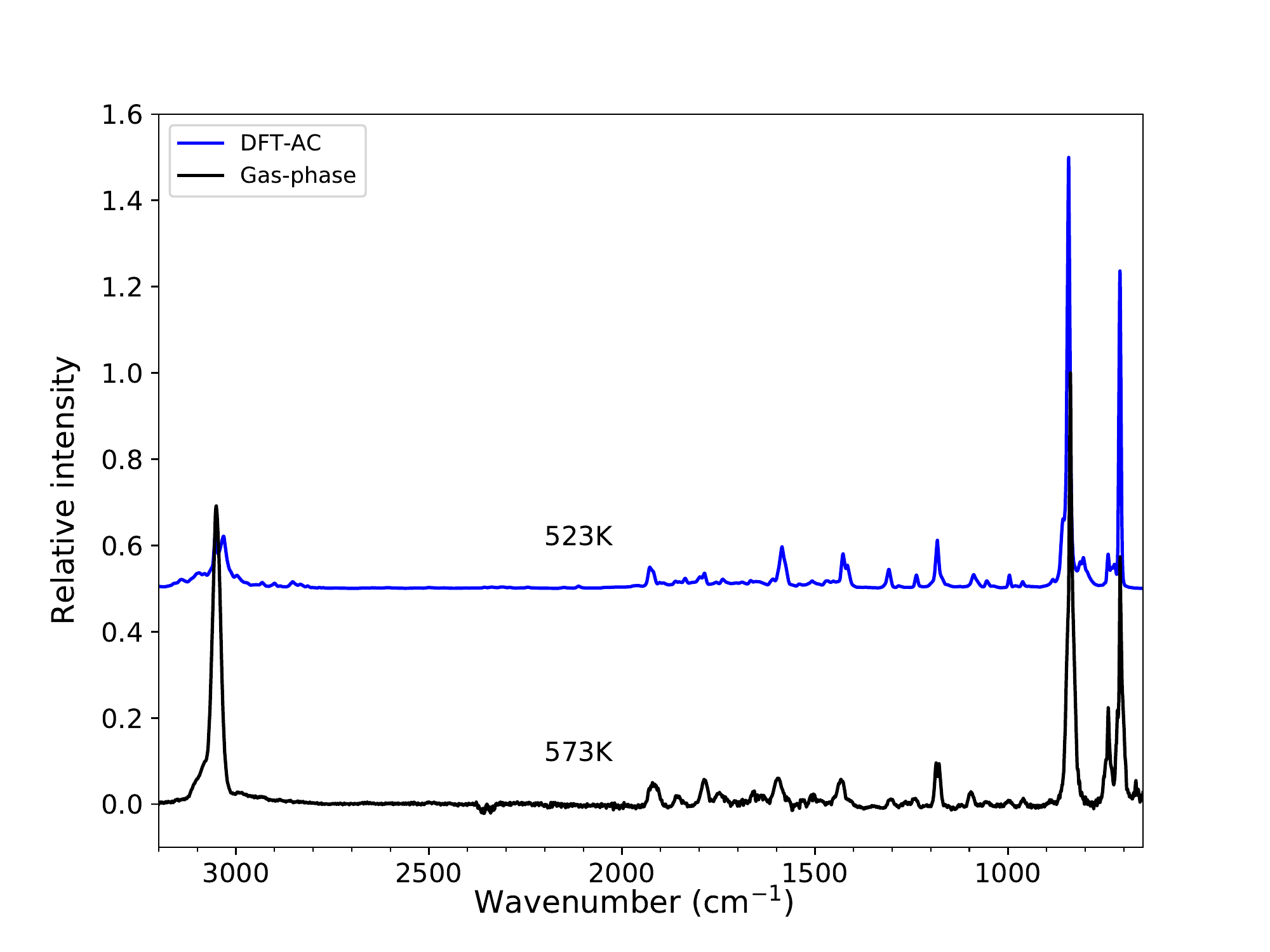} \\
\caption{Top: theoretical infrared spectrum of pyrene from 3200 to 650 \cm at 300\,K, simulated using DFT-AC together with the experimental \cite{chakraborty2019} condensed phase spectrum at 300K; bottom: theoretical infrared spectrum of pyrene from 3200 to 650 \cm at 523\,K, simulated using DFT-AC together with the experimental \cite{Joblin1995} gas-phase spectrum at 573\,K. In all cases the spectra were normalized to unit peak intensity of the most intense band of pyrene at 844.4 \cm. Theoretical spectra computed by AC at 300 and 523\,K have been vertically shifted compared to the solid and gas-phase experimental spectra for clarity.}
\label{fig:anhcaos_vs_exp}
\end{figure*}

\begin{figure*}
    \centering
    \includegraphics[width=14cm]{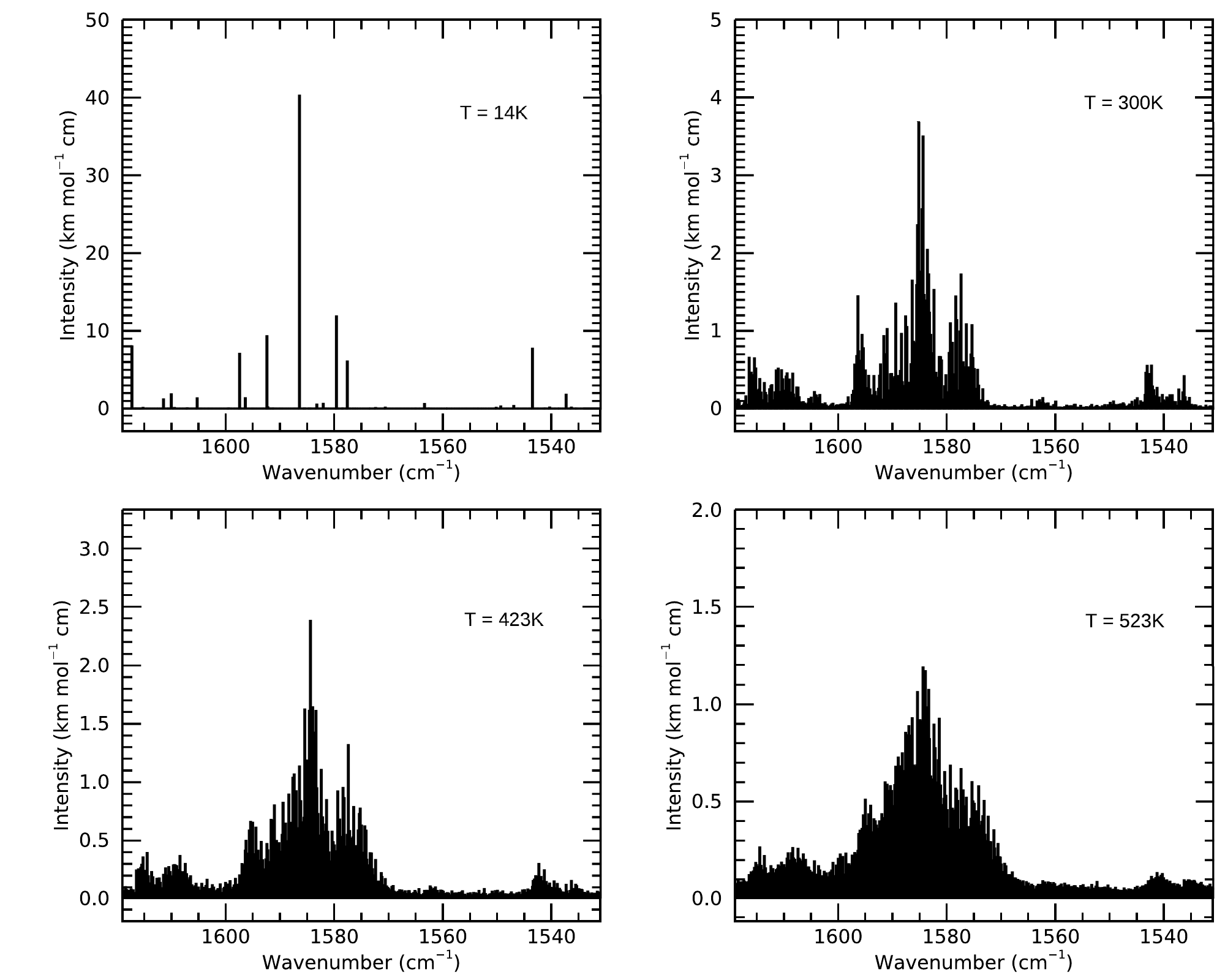}
    \caption{Stick spectra of pyrene from 1620 to 1530 cm$^{-1}$ at 14, 300, 423 and 523\,K computed using DFT-AC.}
    \label{fig:ac_stick1}
\end{figure*}

\begin{figure*}
    \centering
    \includegraphics[width=14cm]{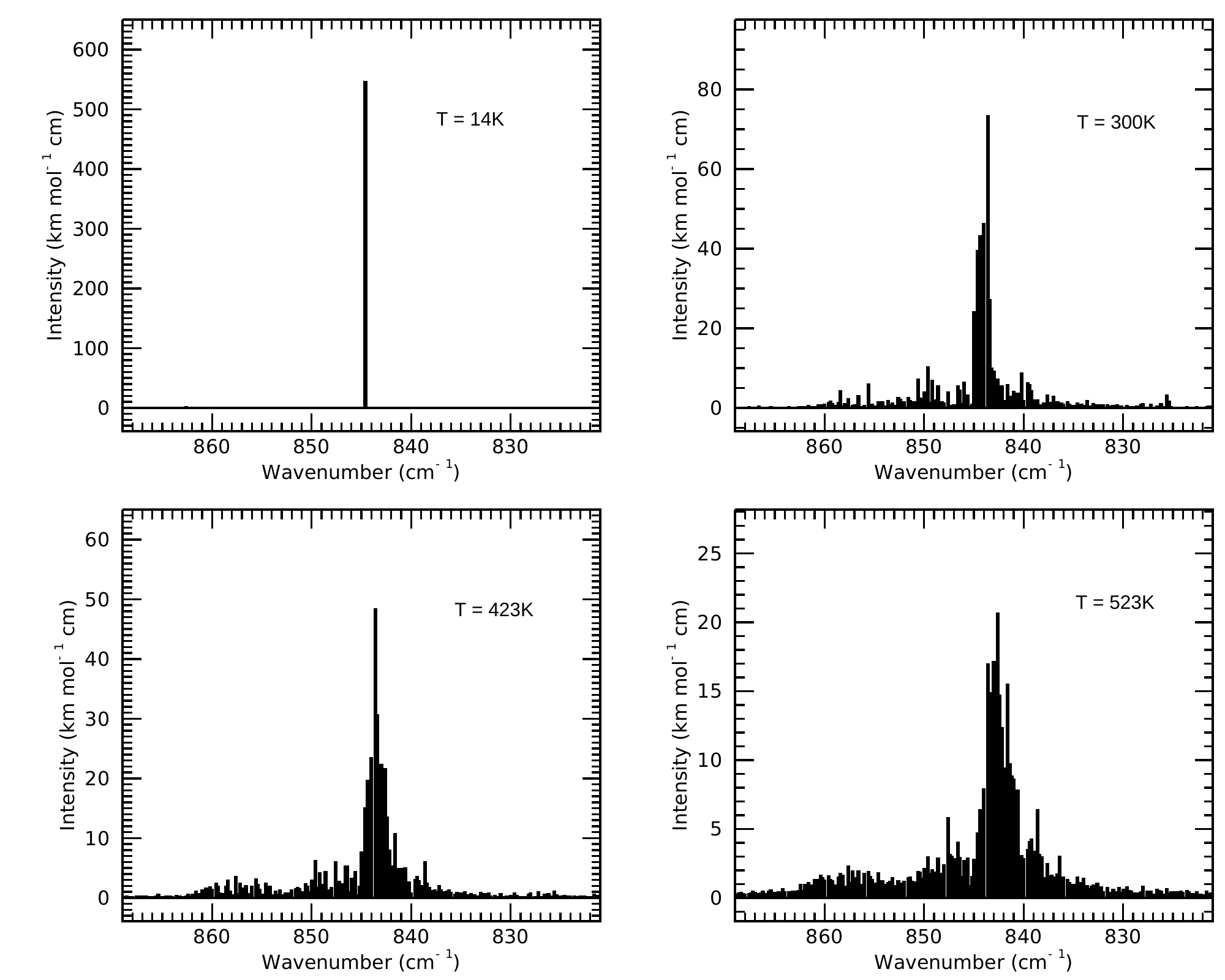}
    \caption{Stick spectra of pyrene from 870 to 820 cm$^{-1}$ at 14, 300, 423 and 523\,K computed using DFT-AC.}
    \label{fig:ac_stick2}
\end{figure*}

\begin{figure*}
    \centering
    \includegraphics[width=14cm]{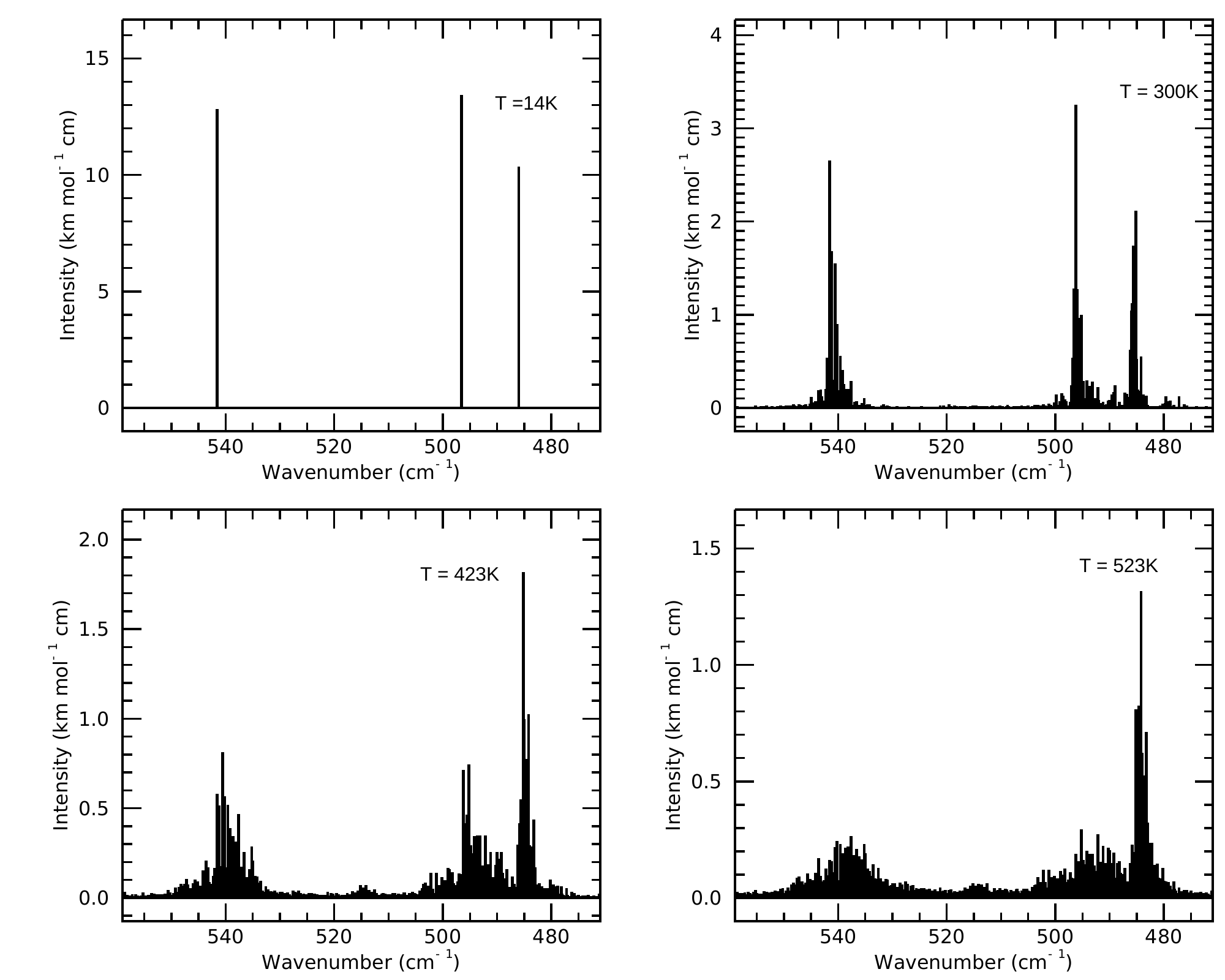}
    \caption{Stick spectra of pyrene from 470 to 560 cm$^{-1}$ at 14, 300, 423 and 523\,K computed using DFT-AC.}
    \label{fig:ac_stick3}
\end{figure*}

\begin{figure*}
     \centering
     \includegraphics[width=14cm]{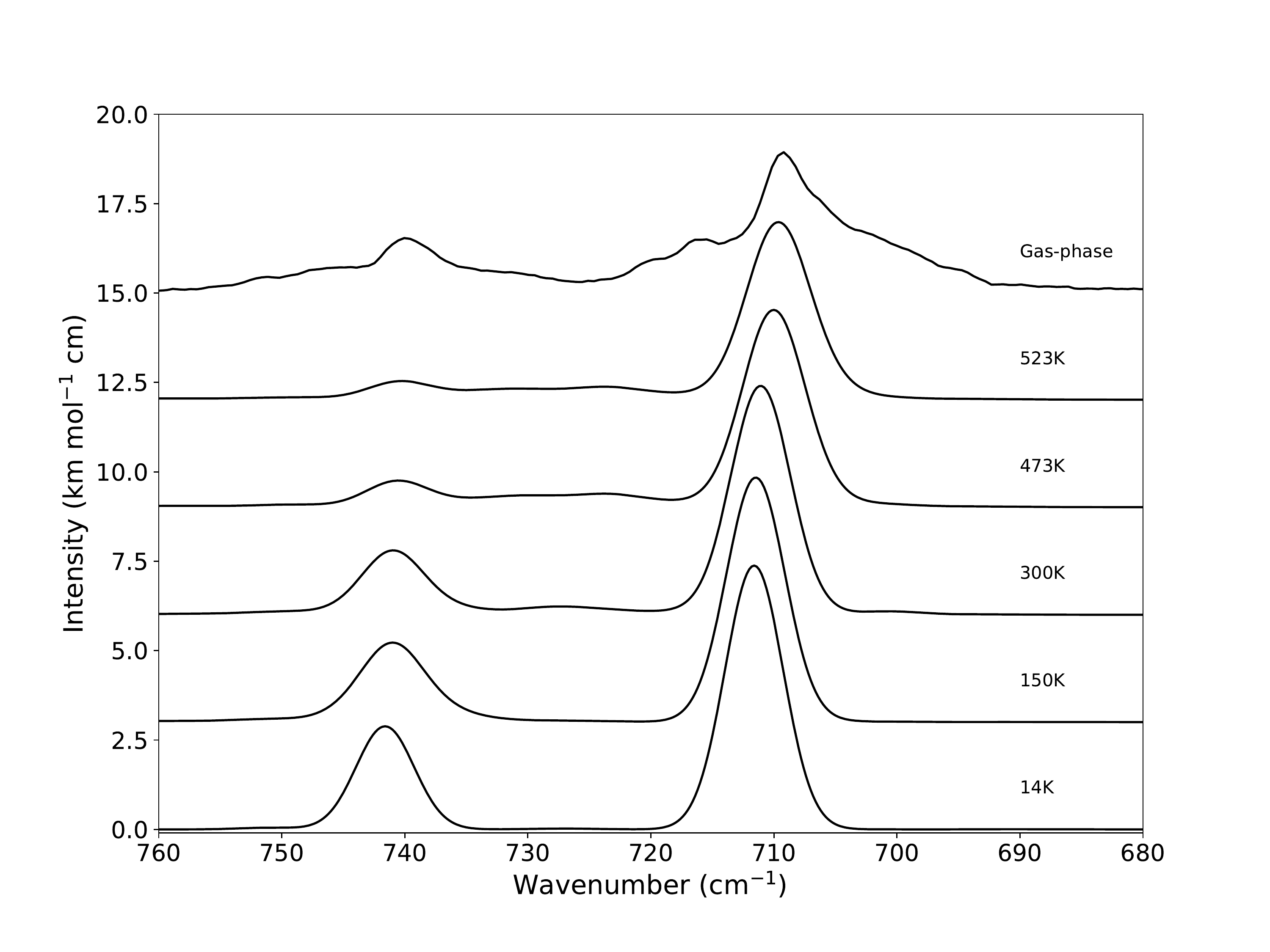}
     \caption{Temperature evolution of the band profiles for two bands of pyrene at 741.1 and 711.6 \cm simulated  using DFT-AC method together with gas-phase data at 573\,K\cite{joblin1994}.}
     \label{fig:741_711_band_profile}
 \end{figure*}

\begin{figure*}
     \centering
     \includegraphics[width=14cm]{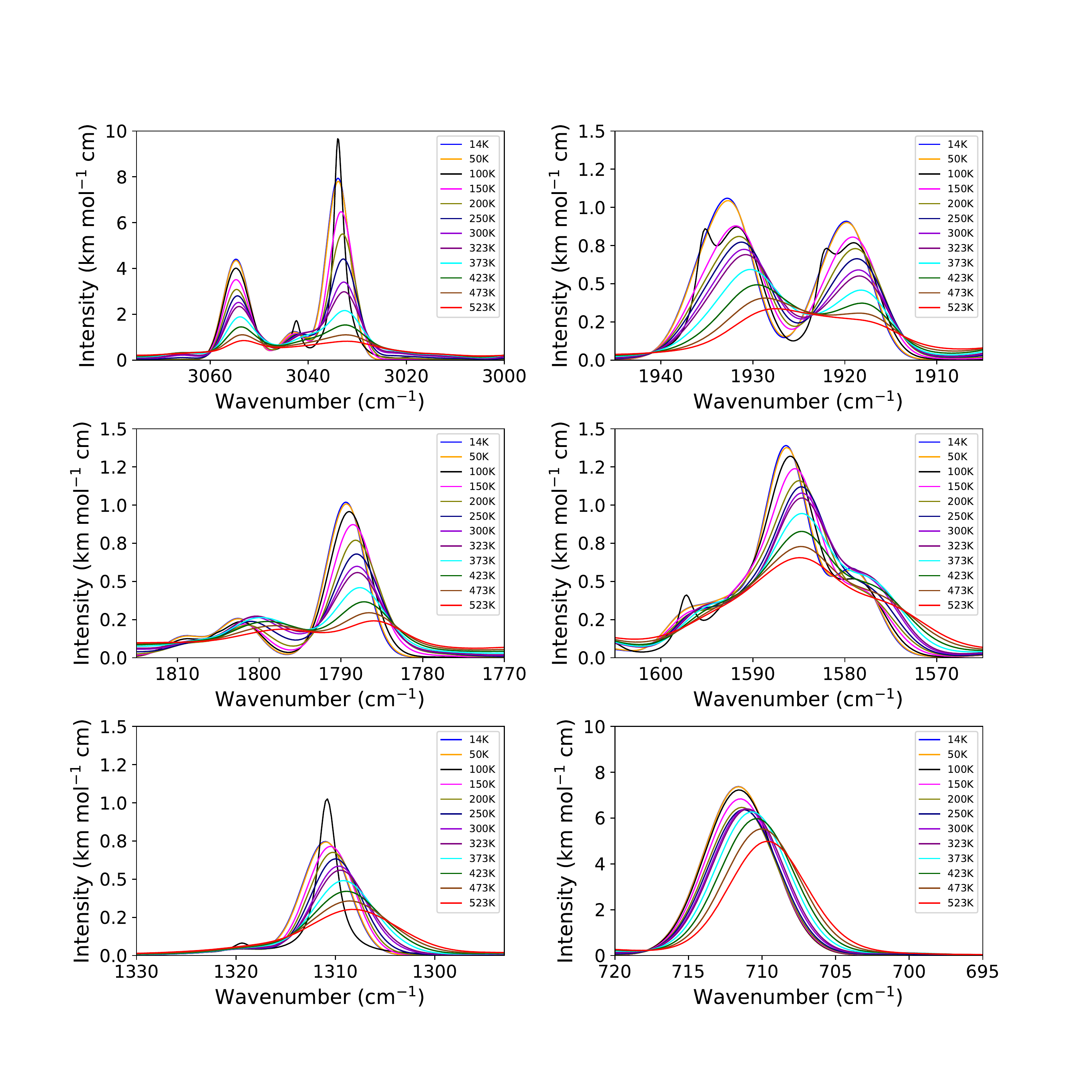}
     \caption{Temperature evolution of the simulated band profiles for a selection of bands of pyrene simulated  using DFT-AC method.}
     \label{fig:pyrene_band_profile}
 \end{figure*}

The raw spectra resulting from AnharmoniCaOs are ``stick'' spectra, each ``stick'' resulting from an individual anharmonic transition and arbitrarily (in principle infinitely) sharp, since AnharmoniCaOs does not include rotational structure. For the results presented here, the resolution of the calculations yields a ``stick'' width of 0.2~cm$^{-1}$. Figure~\ref{fig:anhcaos_vs_exp} gives an overall comparison with available experimental absorption spectra, both in solid state and in gas-phase \cite{chakraborty2019,joblin1994}. For viewing convenience, and easier comparison with experimental data, in which bands are broadened either by unresolved rotational structure (in gas-phase) or by solid state effects, we convolved DFT-AC spectra with Gaussian profiles with a FWHM of $\sim$5~cm$^{-1}$. Figure~\ref{fig:anhcaos_vs_exp} compares theoretical DFT-AC spectra with available experimental spectra at 300 and 523 K, all being normalised to the maximum peak intensity.
At this large scale, the agreement between both types of spectra appears very good both for positions and relative intensities. This applies to fundamentals but also to combination and overtone bands, permitting their individual identification with fairly high accuracy. \\
Figure~\ref{fig:ac_stick1} zooms in on the spectral region around the band at $\sim$1583~cm$^{-1}$ and shows the DFT-AC absorption spectrum computed at temperatures ranging from 14~K to  523~K, in standard absolute molar absorption intensity units (km mol$^{-1}$ cm). Beside the fundamental band near 1583~cm$^{-1}$, even at low temperature one can clearly see many additional combination bands. Relevant individual bands in this spectral range were identified in \citet{Mulas2018}. They can be rather accurately predicted thanks to the inclusion in AnharmoniCaOs of the second derivatives of the dipole moment, and to the detailed treatment of resonances. On the other hand, this detailed treatment of resonances, which in case of several terms of the dipole moment expansion contributing to the same transition causes the code to merge several polyads in a single, larger one, renders the calculation increasingly heavy at higher excitation energies, making it computationally expensive, while still feasible, at very high temperatures. With rising temperature, a huge number of additional hot bands progressively appear, corresponding to transitions arising from increasingly excited states. These cluster mostly around the individual bands already present at low T, with variable shifts around them, resulting in a clear broadening and, in some cases, shift of band centroids. Bands which are close at low T tend to blend at high T, making it difficult, or even impossible, to resolve them beyond some temperature. In addition, a multitude of very weak, resonance-activated features appear, with increasing T, even farther away from the main bands, correspondingly causing a reduction in the intensity of the latter. This results, qualitatively, in main bands losing a fraction of their intensity at high T, which goes into a featureless, broad plateau below them. In theoretical spectra, this intensity is still present in the spectrum, spread over a wider spectral range; in experimental ones, even if present (and there is no reason why it should not be), it is further spread by rotational structure or solid-state effects, it thus merges with the continuum, cannot be clearly distinguished from it, and is thus easily partially or completely subtracted away with it, being lost to measurement.\\
Figure~\ref{fig:ac_stick2} zooms in on the band at $\sim$844~cm$^{-1}$, for an example of a spectral region with a well-resolved, single fundamental band. With the exception of the blending of initially resolved features, the same behaviour described above is observed, with hot bands appearing all around the fundamental, producing a shift of the centroid of the resulting band, as well as a broadening. In this case, as well, we also observe the appearance, at the highest T, of a ``carpet'' of very weak bands, ``stealing'' some of the intensity of the fundamental and spreading it over a broad plateau.\\
Figure~\ref{fig:ac_stick3} gives another example to illustrate the remarkably large variability of bandwidths with temperature. In this case, one band remains relatively narrow with increasing T, while neighbouring ones show a much larger broadening. \\
Figure~\ref{fig:741_711_band_profile} shows the evolution of the band profile for some out of plane CH bending modes, as predicted by DFT-AC for T ranging from 14~K to 523~K, compared with the gas-phase spectrum at the closest temperature, i.~e. 573~K. The DFT-AC spectra in this figure were convolved with a Gaussian as for Figure~\ref{fig:anhcaos_vs_exp}.
All absorption spectra in this figure are shown in absolute units (km mol$^{-1}$ cm). The band positions predicted by DFT-AC very clearly tend to the ones in the gas-phase spectrum, which is broadened by rotational structure not included in AnharmoniCaOs. The peak intensity predicted for the band at $\sim$741~cm$^{-1}$ clearly decreases with temperature, as it is distributed over many hot bands around it and many more, weakly resonant features further away.\\
Figure~\ref{fig:pyrene_band_profile} shows some more bands simulated by DFT-AC for the same temperature range. Again, these spectra were convolved with a Gaussian as for Figures~\ref{fig:anhcaos_vs_exp} and \ref{fig:741_711_band_profile}. They are shown in absolute units (km mol$^{-1}$ cm). Many of the bands present significant resolved structure at low temperature, which is progressively lost as T increases. All of them show the general behaviour we described above.\\

\begin{table*}
\begin{tabular*}{\hsize}{@{\extracolsep{\fill}}cccccc}
\hline
Gross pos.($\mu$m)  & Integration range (\cm)  & \multicolumn{4}{c}{Integrated intensities (\km)}  \\
\cline{3-6}
  & & Gas 570\,K\cite{joblin1994}  &  \multicolumn{3}{c}{DFT-AC} \\
 \cline{3-6}
& &  &  0\,K\cite{Mulas2018} & \multicolumn{1}{c}{300\,K} & 523\,K\\ 
\cline{4-6}
\hline
3.3	 & [2850-3250] & 140($\pm$10)   & 99 &78.0 & 62.0 \\
5.2  & [1900-1950] & 9.5 ($\pm$0.5)   &  9.0 &12.3  & 7.6 \\
5.3	 & [1830-1880] & 3.8 ($\pm$0.1)   & 5.5 & 4.8  & 5.0 \\
5.6  & [1780-1830] & 7.8 ($\pm$0.7)   &  8.8 &8.7  &6.9  \\
5.7  & [1715-1760] & 5.8 ($\pm$0.2)   &  2.7 &5.0  &4.5 \\
5.9  & [1677-1704]    & --		      & 1.9 &2.6  & 2.4\\
6.0	 & [1655-1677] & 2.5 ($\pm$0.2)   & 2.7 &2.2  & 2.2\\
6.1	 & [1630-1655] & --   & 2.8 &2.5  & 2.5\\
6.2  & [1560-1620] & 11.4 ($\pm$1.2)  & 17.3 &18.3  &15.4  \\
6.6  & [1531-1552] & --		     &  2.6 & 1.7 &  1.2\\
6.8  & [1467-1525] & --		     & 3.7 & 6.6 & 4.9 \\
7.0  & [1405-1430] & 11.4 ($\pm$0.1)   & 8.8 &10.1  & 8.5 \\
7.7	 & [1300-1327] & 2.4 ($\pm$0.1)  & 5.2 &  4.4& 3.8 \\
8.0	 & [1224-1245] & 1.9 ($\pm$0.3)    & 2.6 &2.4  & 2.1 \\
8.4	 & [1165-1200] & 10.5 ($\pm$0.1)   & 13.7 & 13.8 & 10.2 \\
9.1	 & [1074-1100] & 5.4 ($\pm$1.2)	   & 5.4 & 5.7 & 3.5 \\
10.0 & [988-1003]  & 2.2 ($\pm$0.6)      &1.7   & 1.6  & 1.6 \\
11.8 & [824-863]   & 100 ($\pm$6) 	   & 101 & 103.6 & 77.9 \\
13.4 & [730-750]   & 20.8 ($\pm$1.4)  	  & 15.2 & 13.2 &5.9 \\
14.0 & [695-725]   & 46 ($\pm$1) 	   & 45.3 & 40.7  & 36.7 \\
18.5 & [526-550]   &              & 2.3 & 2.3 & 1.4 \\
20.0 & [487-501]   &               & 2.7	&3.0  &1.6\\
20.5 & [474-487]   &	 	 	   & 1.9 & 1.5 & 1.9\\
29.0 & [335-354]   &	 	 	    & 1.4 &1.6  &1.5 \\
50.0 & [185-212]   &	 	 	    & 9.7 &10.9 & 5.7\\

\hline
\end{tabular*}
\caption{Integrated intensities of the bands computed using DFT-AC at 0, 300 and 523\,K together with the gas-phase experimental ones}\label{tab:pyrene2}
\end{table*}

\subsection{DFTB Molecular dynamics results}\label{sec:dftbmd_results}

The DFTB spectrum computed at the harmonic level at 0\,K is a stick spectrum, whereas the spectra obtained at finite temperature from DFTB-MD simulations present broadenings, shifts and merging of bands. The latters appear naturally, incorporating the anharmonic effects of the full DFTB PES without truncation of the latter at a given order. Thermal effects result from MD explorations, and not from an \textit{a posteriori} convolution of these spectra with gaussian or lorentzian functions. One should however remember that the frequency resolution scales as the reciprocal of the total duration of the MD segments used to numerically evaluate equation \ref{Eq:specdyn}. In this work the frequency resolution is 1.7~cm$^{-1}$. 

\begin{figure*}
    \centering
    \includegraphics[width=14cm]{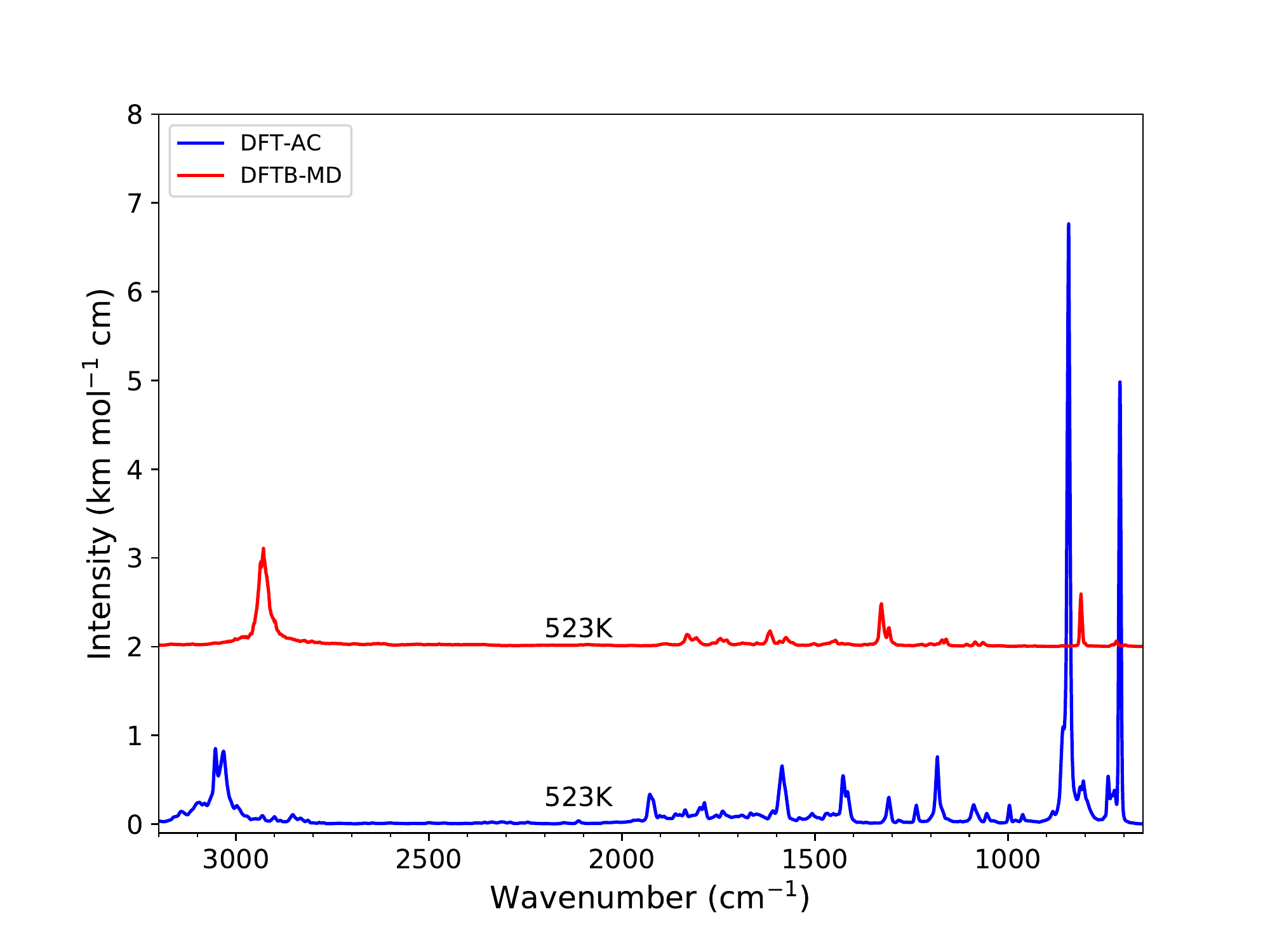}
    \caption{Theoretical infrared spectrum of pyrene from 3200 to 650 \cm at 523\,K, simulated using DFT-AC and DFTB-MD. The DFTB-MD spectrum is vertically shifted for clarity by 2~km~mol$^{-1}$~cm as in Figure~\ref{fig:anhcaos_vs_exp}.}
    \label{fig:pyrene_anh_demon}
\end{figure*}

As an example, the spectrum computed at 523\,K is shown on Figure \ref{fig:pyrene_anh_demon} together with the highest temperature spectrum (523\,K) obtained by DFT-AC. As a first look, the two spectra present the following common features: a wide band around 3000~cm$^{-1}$;  two well defined bands below 1000~cm$^{-1}$ and a forest of bands between 1000 and 2000~cm$^{-1}$. The main differences rely on the band intensities, in particular for the DFTB bands at 3000 and 720~cm$^{-1}$ which have, respectively, higher and lower intensities when compared to their equivalent in the DFT-AC model. We remark here that, with respect to bands positions, relative bands intensities are much more challenging to reproduce by an approximated model and this remains true for \textit{ab initio} schemes. As a consequence, it is more difficult to reproduce the spectral evolution of a specific band located in a region of high vibrational states density. In such a region, when the temperature increases, many bands will blend in a single, unresolved spectral feature. The large errors, due to DFTB, in the estimated intensities of individual, unresolved components will then severely affect the predicted evolution of the profile and position of the resulting blended feature.

\begin{figure*}
     \centering
     \includegraphics[width=14cm]{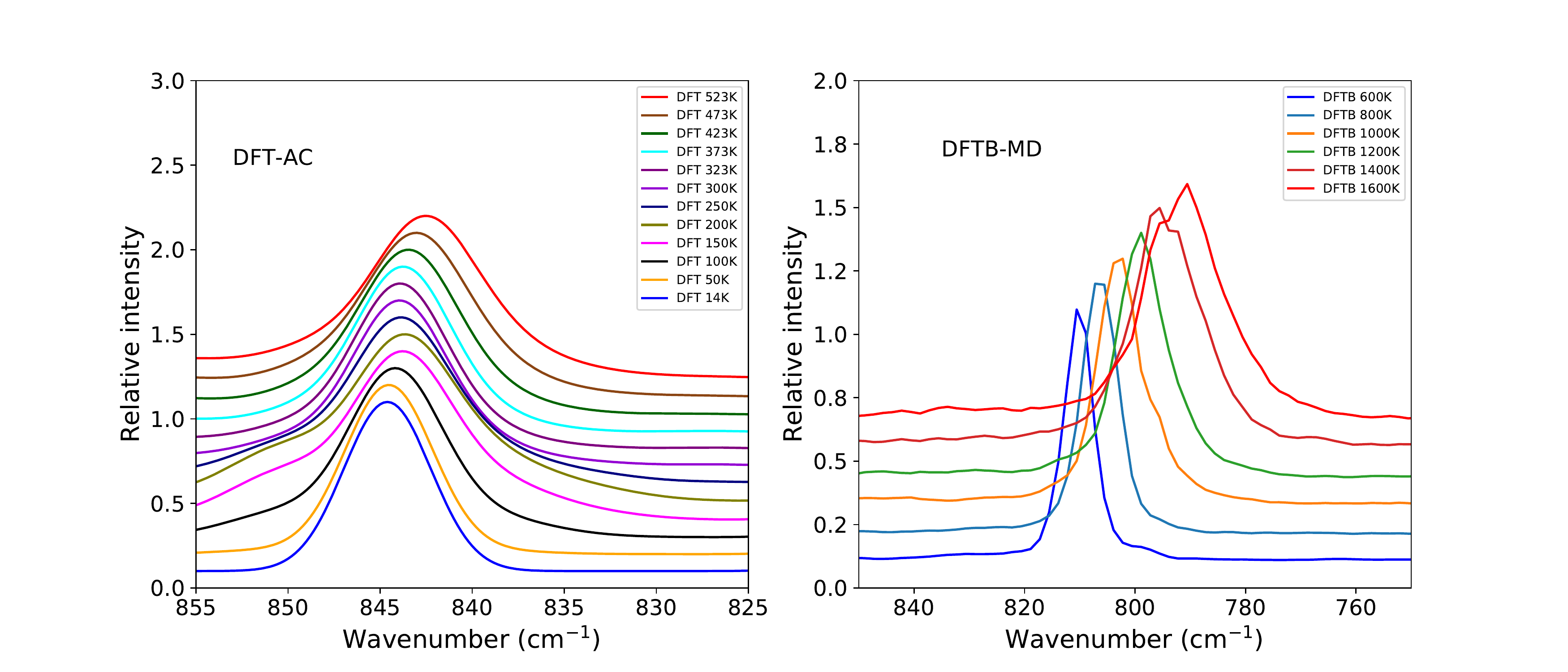}
     \caption{Temperature evolution of the highest intensity band of pyrene at 844.4 cm$^{-1}$ (cf. Table \ref{Tab:anh_shift}) simulated using DFT-AC and DFTB-MD. Intensities are given in arbitrary units, normalised to unit peak intensity, to emphasize the evolution of the spectral profile.}
     \label{fig:pyrene_11_2_profile}
\end{figure*}

DFT and DFTB spectra computed at the harmonic levels allowed us to identify the nature of the bands on the basis of vibrational modes visualisation, the DFT and DFTB normal modes being similar. Table S3 in Supplementary information lists all harmonic normal modes resulting from the DFT vs DFTB calculations, matching all of them after visual inspection of the corresponding nuclear motions. This is an intermediate step to determine which band at the DFT-AC level can be compared with which band at the DFTB-MD level, but the actual comparison hereafter is never done at this harmonic level. Since some bands, in DFTB-MD spectra, become so shallow to be hardly distinguishable from the continuum, making the measurement of their position and width unreliable, we focus here on the evolution of band position and width for four relevant bands: bands~1 and 11 which are the two most intense ones; band 14 which is a satellite of band~11 whose evolution with temperature allowed for unambiguous derivation of band position and width and, finally, band~2 as representative of a more complex spectral region. The band positions for these 4 bands at 0\,K (harmonic) and  600\,K are reported in table \ref{Tab:anh_shift}. When compared to experimental values and keeping in mind the level of theory, the DFTB-MD band positions at 600\,K are in relatively good agreement with experimental results for bands~1, 3 and 14, presenting relative errors of 5.1 , 3.1 and 2.6~\%, a larger error of 15~\% being observed for band~2. 
It can also be seen that the differences between DFT and DFTB band positions for temperatures around 600\,K are already present at the harmonic level: bands~1, 11 and 13 being redshifted in the DFTB spectra with respect to DFT spectra, the opposite being observed for band~2. Figure~\ref{fig:pyrene_11_2_profile} shows the evolution with temperature of the most intense band, namely band~11,  at the DFT-AC and DFTB-MD levels. It shows that, despite the fact that the absolute band positions differ between the two models, the trend, i.e. a redshift and broadening when increasing the temperature, seems to be consistent, encouraging the use of the DFTB-MD model to compute the bands \textit{evolution} at high temperatures rather than their  absolute positions.

\begin{table*}
\centering
\footnotesize
\caption{Calculated band positions of four representative IR bands of pyrene. Experimental values are listed from the work of \citet{chakraborty2019} for the low temperature part and \citet{joblin1994} for the high temperature measurements}

\hspace{2cm}
\begin{tabular}{c c c c c |c c |c c}
\multicolumn{1}{c}{}&\multicolumn{6}{c|}{Theory}&\multicolumn{2}{c}{Experiment}\\
\hline 
\multicolumn{1}{c}{}&\multicolumn{4}{c}{DFT}& \multicolumn{2}{|c|}{DFTB}& \multicolumn{2}{c}{ }\\
\multicolumn{1}{c}{}&\multicolumn{1}{c}{Harmonic}&\multicolumn{3}{c|}{AnharmoniCaOs}&\multicolumn{1}{|c}{Harmonic}&\multicolumn{1}{c|}{MD}&\multicolumn{1}{c}{Solid\cite{chakraborty2019}}&\multicolumn{1}{c}{Gas\cite{joblin1994}}\\
Temp &0\,K & 0\,K & 300\,K & 523\,K &0\,K& 600\,K &300\,K &570\,K\\
\hline
Band1 &3164.6 & 3040.0& 3040.9 & 3040.5& 2960.3&2897.3& - &3052.0\\
Band2 &1635.4 & 1600.0 & 1585.3 & 1585.8& 1856.3& 1842.2& - &1597.0\\
Band11 & 860.5& 843.0& 844.4& 843.9& 818.4& 814.1&840.0 &840.0\\
Band13 &755.6 & 739.0 & 741.1 & 739.7& 725.1&721.7& 749.4& 740.0\\
\hline
\label{Tab:anh_shift}
\end{tabular}
\end{table*}

\subsection{Anharmonic parameters}\label{sec:anhparameters}
Analysing the DFT-AC spectra as described in Sect.~\ref{sec:analysis}, we identified bands, or band groups, which could be followed for the whole temperature range spanned by DFT-AC calculations, and determined their T-dependent centroids and widths.
The same procedure was applied for the DFTB-MD spectra.

\begin{figure*}
    \centering
    \includegraphics[width=13.5cm]{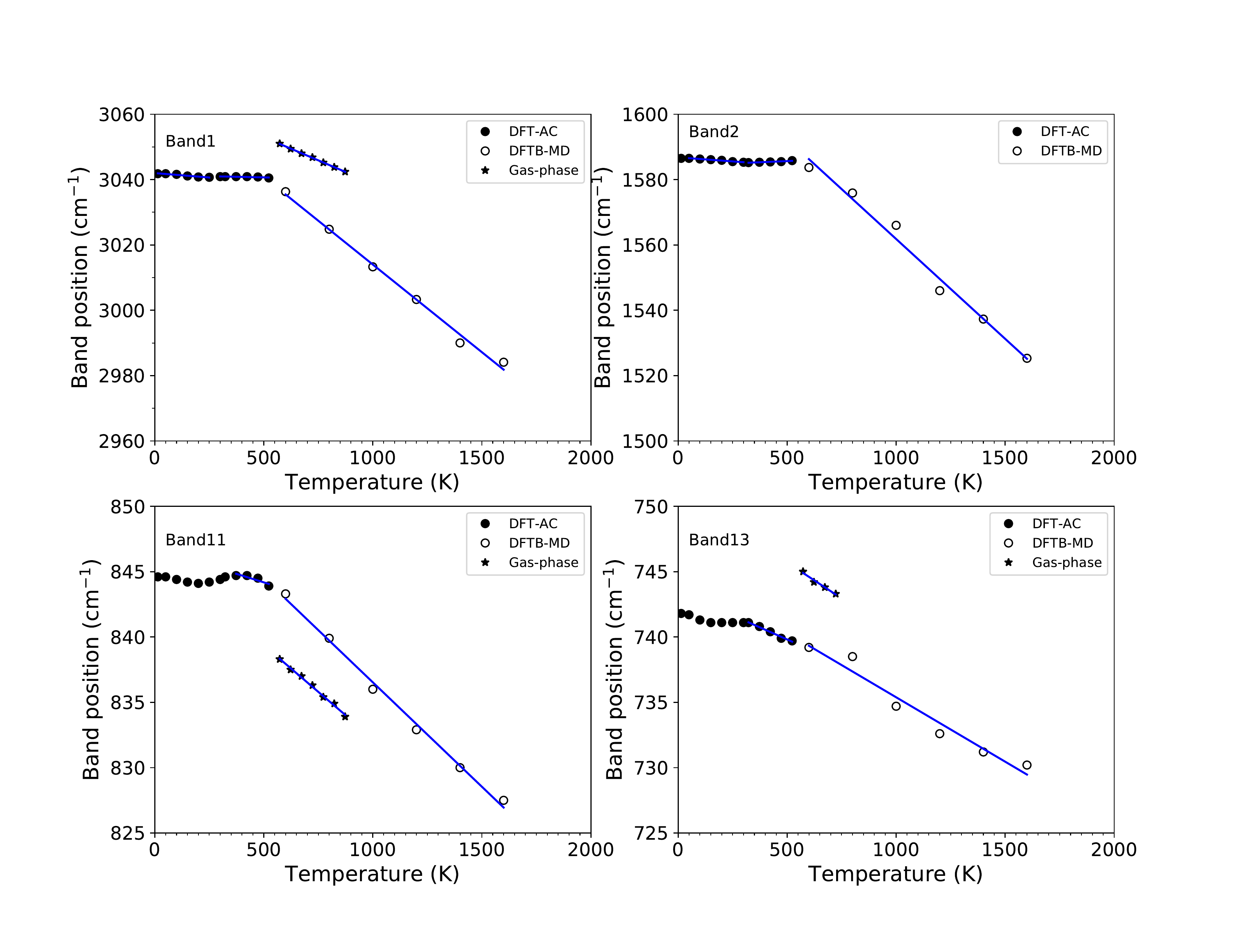}
    \caption{Evolution of band positions with temperature obtained from DFT-AC (filled circles), DFTB-MD (empty circles) calculations and gas-phase experimental from \citet{Joblin1995} (black stars). DFTB-MD data points have been rigidly shifted to smoothly join DFT-AC ones near $\sim$550\,K. These shifts are respectively  +139, -230, +34 and +17.5~cm$^{-1}$ for band~1, band~2, band~11, and band~13. Bands are labelled following Table \ref{Tab:anh_shift}. For band~13 the gas-phase band positions overlaps with DFTB-MD band positions therefore we have shifted all gas-phase data by +5~cm$^{-1}$ for clarity. }
    \label{fig:bp_T_anh_demon_1}
\end{figure*}

Figure~\ref{fig:bp_T_anh_demon_1} reports together the DFT-AC and DFTB-MD positions for the four specific bands selected in Section~\ref{sec:dftbmd_results}. DFT-AC and DFTB-MD positions for all other bands are given in Figures S1 and S2 in the Supplementary Information, respectively. DFTB-MD positions in Figure~\ref{fig:bp_T_anh_demon_1} are rigidly shifted to smoothly join DFT-AC values at 523~K. Three panels in the same figure also report data from available laboratory measurements \citep{Joblin1995}.
Table~\ref{Tab:anh_fact} reports the slopes $\chi'$ of the evolution of band positions vs. temperature for T ranges in which they behave linearly.

\begin{figure*}
    \centering
    \includegraphics[width=13.5cm]{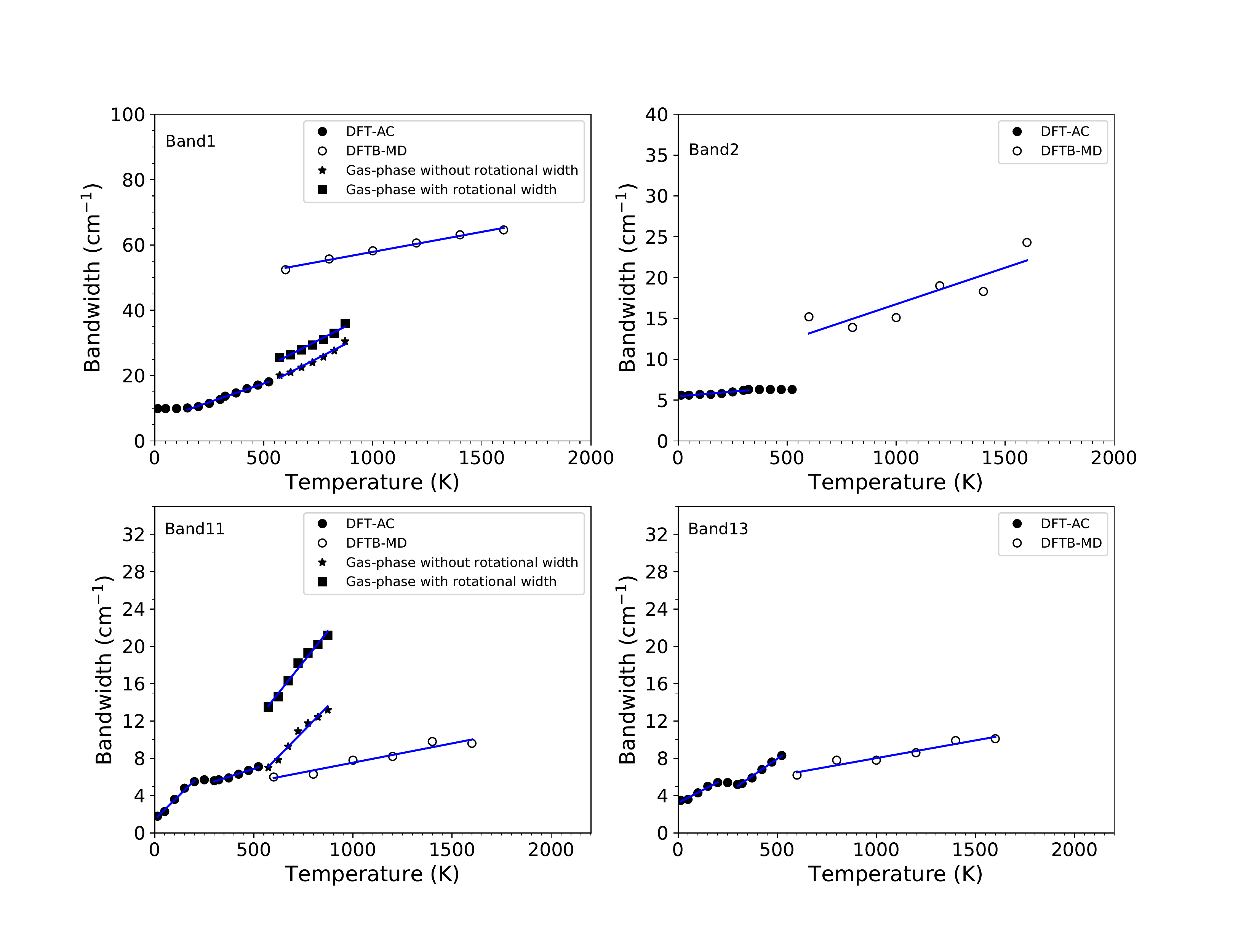}
    \caption{Evolution of bandwidths with temperature obtained from DFT-AC (filled circles), DFTB-MD (empty circles) calculations. Bandwidths from laboratory data \citep{Joblin1995} are shown as they are (filled squares), and after subtraction of the estimated contribution due to molecular rotation in thermal equilibrium (black stars, see text for details). Bands are labelled following Table \ref{Tab:anh_shift}  }
    \label{fig:bw_T_anh_demon_2}
\end{figure*}

Similarly, Figure~\ref{fig:bw_T_anh_demon_2} shows bandwidths determined from DFT-AC and DFTB-MD for the four selected bands, others being given in Figures~S3 and~S4 of the Supplementary Information. Laboratory measurements are included for three bands \citep{Joblin1995}. Since the latter were performed in gas-phase in thermal equilibrium, they also include a significant contribution from rotational structure, which is included neither in DFT-AC nor in DFTB-MD modelling. We therefore estimated the rotational contribution to bandwidth, computing rotational profiles for pyrene at the temperatures of the laboratory data points. For our purposes, we used a simple rigid rotor model, with rotational constants taken from \citet{Malloci2007}. Indeed, from high resolution vibrational spectra, with resolved rotational structure, it appears that variations of rotational constants of pyrene with vibrational quantum numbers, as well as due to centrifugal distortion, are negligible \citep{brumfield2012}. To estimate the rotational width, we applied Equations~\ref{eq:bandcentroid} and \ref{eq:bandwidth} given in Section~\ref{sec:analysis} to the simulated rotational envelopes of a-, b-, and c-type pyrene bands (pyrene is an asymmetric rotor). The rotational contribution to bandwidth is estimated to range from $\sim$6.5~cm$^{-1}$ at 573\,K to $\sim$8~cm$^{-1}$ at 873\,K for band~11 (an out of plane vibrational mode, fully c-type), and from $\sim$5.4~cm$^{-1}$ at 573\,K to $\sim$6.7~cm$^{-1}$ at 873\,K for band 1 (in plane C-H stretches, with similar contribution from a- and b-type bands). To estimate the contribution of vibrational anharmonicity to the bandwidths measured in gas-phase, we followed the approach of \citet{Pech2002}, subtracting the estimated rotational contribution. The resulting ``corrected'' laboratory data points are reported in Figure~\ref{fig:bw_T_anh_demon_2}. 
No shift was applied to DFTB-MD data points in Figure~\ref{fig:bw_T_anh_demon_2}.  Table~\ref{Tab:anh_fact_chi_dp} shows the slopes $\chi''$ of bandwidths, similarly to Table~\ref{Tab:anh_fact} for band positions. \\

\begin{table*}
\centering
\footnotesize
\caption{Empirical anharmonicity factors ($\chi'$ in 10$^{-2}$\cm\K) of the major IR bands of pyrene derived from the linear fitting of the band positions in different temperature ranges (Fig.\ref{fig:bp_T_anh_demon_1} and Figs. S1 and S2 in supplementary material) calculated using DFT-AC and DFTB-MD methods. Experimental values are listed from the condensed-phase measurements by \citet{chakraborty2019} and the gas-phase measurements by \citet{Joblin1995}}
\hspace{2cm}
\resizebox{\textwidth}{!}{\begin{tabular}{c c c c c c c c c c c}
\hline
\multicolumn{1}{c}{No}&\multicolumn{2}{c}{}&\multicolumn{4}{c}{Theory}&\multicolumn{4}{c}{Expt.}\\
\cline{4-7} 
\multicolumn{1}{c}{}&\multicolumn{2}{c}{Position}&\multicolumn{2}{c}{AC}&\multicolumn{2}{c}{MD}&\multicolumn{2}{c}{Solid}&\multicolumn{2}{c}{Gas}\\
\multicolumn{1}{c}{}&\multicolumn{1}{c}{$\mu$m}&\multicolumn{1}c{}{\cm}&\multicolumn{1}{c}{Fit range}&\multicolumn{1}{c}{$\chi'_\mathrm{AC}$}&\multicolumn{1}{c}{Fit range}&\multicolumn{1}{c}{$\chi'_\mathrm{MD}$}&\multicolumn{1}{c}{Fit range}&\multicolumn{1}{c}{$\chi'_\mathrm{rec}$\cite{chakraborty2019}}&\multicolumn{1}{c}{Fit range}&\multicolumn{1}{c}{$\chi'_\mathrm{gas}$}\\
\multicolumn{
1}{c}{}&\multicolumn{2}{c}{}&\multicolumn{1}{c}{(K)}&\multicolumn{1}{c}{}&\multicolumn{1}{c}{(K)}&\multicolumn{1}{c}{}&\multicolumn{1}{c}{(K)}&\multicolumn{1}{c}{}&\multicolumn{1}{c}{(K)}&\multicolumn{1}{c}{}\\
\hline
1&3.3  & 3040.9& 300 - 523 &$-$0.1& 600 - 1600 & $-$6.4&- &- & 573 - 873&$-$2.5\\
2&6.3 & 1585.3 &373 - 523 & $-$0.3&600 - 1600 &$-$6.8 &- &- & & \\
3&6.8  & 1472.4 & 300 - 523 & $-$1.2 & 600 - 1600 & $-$5.2 &300 - 723 &$-$1.3  & &\\
4& 7.0  & 1428.9& 300 - 523 & $-$0.2& 600 - 1600 & $-$5.2 &300 - 723 &$-$1.4&573 - 673 &$-$1.1\\
5& 7.6  & 1309.9& 300 - 523 & $-$0.3& 600  - 1600 & $-$4.2 &200 - 723 &$-$1.3 $\pm$ 0.25& &\\
6&8.1  & 1239.2& 250 - 523 & $-$1.1 & 600 - 1600 & $-$1.5 & 14 - 723&$-$0.4$\pm$0.4 & &\\
7& 8.5  & 1183.4& 323 - 523 & $-$0.9&- & - & 150 - 723&$-$0.7 $\pm$ 0.15  &573 - 873 &$-$0.9\\
8& 9.2  & 1090.1& 250 - 523 & $-$1.3 & - & - & 150 - 723&$-$0.5 $\pm$ 0.15  &573 - 673 &$-$0.2\\
9& 10.1 & 994.9& 300 - 523 & $-$0.2& - & - & 150 - 723&$-$0.4 $\pm$0.4  & &\\
10& 10.4 & 963.5 & 473 - 523 & $-$0.1& - & - & 150 - 723&$-$1.4  & &\\
11& 11.8&844.4 & 373 - 523 & $-$0.5 & 600 - 1600 & $-$1.6  &150 - 723 &$-$1.1  &573 - 873 &$-$1.4\\
12& 12.2 & 817.8& 300 - 523 & $-$1.6 & - & - & 150 - 723&$-$0.2  & &\\
13& 13.5 & 741.1 & 300 - 523 & $-$0.7 & - & - & 150 - 723&$-$1.5 $\pm$ 0.15  &573-723 &$-$1.1\\
14& 14.1 & 711.6& 300 - 523 & $-$0.4 & 600 - 1600 & $-$0.9 & 150 - 723& $-$0.8  &573-723 &$-$1.1\\
\hline
\label{Tab:anh_fact}
\end{tabular}}
\end{table*}

\begin{table*}
\centering
\footnotesize
\caption{Empirical anharmonicity factors ($\chi''$ in 10$^{-2}$ \cm\K) of the major IR bands of pyrene derived from the linear fitting of the bandwidths in different temperature ranges (Fig.\ref{fig:bw_T_anh_demon_2} and Figs S3 and S4 in supplementary material) calculated using DFT-AC and DFTB-MD methods. For DFT-AC we also report two reference values of the bandwidth at 14 and 300\,K. Experimental values of $\chi''$ are listed from the condensed-phase measurements by \citet{chakraborty2019} and the gas-phase measurements by \citet{Joblin1995}. ``WR'' and ``WOR'' labels the empirical anharmonicity factors respectively as derived from the gas-phase data or after subtraction of the estimated thermal rotational broadening (see text for details). }
\hspace{2cm}
\resizebox{\textwidth}{!}{\begin{tabular}{c c c c c c c c c c c c c c}
\hline
\multicolumn{1}{c}{No}&\multicolumn{2}{c}{}&\multicolumn{6}{c}{Theory}&\multicolumn{5}{c}{Expt.}\\
\cline{4-9} 
\multicolumn{1}{c}{}&\multicolumn{2}{c}{Position}&\multicolumn{4}{c}{AC}&\multicolumn{2}{c}{MD}&\multicolumn{2}{c}{Solid}&\multicolumn{3}{c}{Gas}\\
\multicolumn{1}{c}{}&\multicolumn{1}{c}{$\mu$m}&\multicolumn{1}c{}{\cm}&\multicolumn{2}{c}{Bandwidth}&\multicolumn{1}{c}{Fit range}&\multicolumn{1}{c}{$\chi''_\mathrm{AC}$}&\multicolumn{1}{c}{Fit range}&\multicolumn{1}{c}{$\chi''_\mathrm{MD}$}&\multicolumn{1}{c}{Fit range}&\multicolumn{1}{c}{$\chi''_\mathrm{rec}$\cite{chakraborty2019}}&\multicolumn{1}{c}{Fit range}&\multicolumn{2}{c}{$\chi''_\mathrm{gas}$}\\
\multicolumn{
1}{c}{}&\multicolumn{2}{c}{}&\multicolumn{1}{c}{14\,K}&\multicolumn{1}{c}{300\,K}&\multicolumn{1}{c}{(K)}&\multicolumn{1}{c}{}&\multicolumn{1}{c}{(K)}&\multicolumn{1}{c}{}&\multicolumn{1}{c}{(K)}&\multicolumn{1}{c}{}&\multicolumn{1}{c}{(K)}&\multicolumn{1}{c}{WR}&\multicolumn{1}{c}{WOR}\\
\hline
1 & 3.3  & 3040.9 & 9.9 & 18.1& 300 - 523 & 2.2 & 600 - 1600 & 1.2 & - & - & 573 - 873 & 3.4& 3.4\\
2 & 6.3  & 1585.3 & 5.6 & 6.3& 373 - 523 & 0.2 & 600 - 1600 & 0.9 & - & - & & & \\
3 & 6.8  & 1472.4 & 6.5 & 10.7 & 300 - 523 & 1.5 & 600 - 1600 & 0.3 & 250 - 723 & 1.5  & & &\\
4 & 7.0  & 1428.9 & 4.0 & 6.3&300 - 523 & 0.7 & 600 - 1600 & 0.5 & 300 - 723 & - &  & & \\
5 & 7.6  & 1309.9 & 2.0 & 6.3 &300 - 523 & 1.3 & 600 - 1600 & 0.5 &     -     & - &  & &\\
6 & 8.1  & 1239.2 & 1.3 & 6.9 &250 - 523 & 1.6 & 600 - 1600 & 0.8 & 200 - 723& 0.9 $\pm$ 0.3& & &\\
7 & 8.5  & 1183.4 & 4.2 & 7.6 & 323 - 523 & 1.5 & - &- & 523 - 723& 1.1  &573 - 873 &1.5&1.2\\
8 & 9.2  & 1090.1 & 4.6 & 8.3 & 250 - 523 & 1.4 & - & - & 200 - 723& 1.1 $\pm$ 0.4 &  & &\\
9 & 10.1  & 994.9 & 4.1 & 5.9 &300 - 523 & 0.5 & - & - & 250 - 723& 1.0 $\pm$ 0.4  & & &\\
10 & 10.4 & 963.5 & 1.6 & 9.8 & 473 - 523 & 2.0 & - & - &    &  & & &\\
11 & 11.8 & 844.4 & 1.8 & 7.1 &373 - 523 & 0.7 & 600 - 1600 & 0.4  & &  & 573 - 873 &2.6 & 2.1\\
12 & 12.2 & 817.8 & 3.4 & 8.3 & 300 - 523 & 2.0 & -          &  -   & &  & & &\\
13 & 13.5 & 741.1 & 3.5 & 8.3 & 300 - 523 & 1.5 & -          &  -   & &  & & &\\
14 & 14.1 & 711.6 & 1.7 & 5.1 & 300 - 523 & 0.5 & 600 - 1600 & 0.4  & &   & & &\\
\hline
\label{Tab:anh_fact_chi_dp}
\end{tabular}}
\end{table*}

\section{Discussion}

The DFT-AC approach, privileging accuracy over computational costs, leads to consider larger and larger sets of resonating states, making it less practical and difficult to use on a systematic basis on many species up to very high temperatures. Moreover, the use of truncated Taylor expansions around the equilibrium position for the PES and electric dipole is expected to become less and less accurate at high temperature. The limit of its validity is difficult to estimate since it is mode dependant and complex to define, as it depends on the PES shape.
In the combined scheme presented in this work, the case of high temperatures is treated by decreasing the level of theory from DFT to DFTB in order to enable the computation of IR spectra in a molecular dynamics (MD) scheme capturing all anharmonic effects, i.e. beyond the quartic expansion of the PES. In switching from a fully quantum treatment to classical MD, we lose all information about the discrete structure of hot bands. On the other hand, we gain, in principle, an efficient method that actually gets more efficient with increasing temperatures, since increasing anharmonicity helps to quickly reach ergodic redistribution of vibrational energy. On the other hand, in switching from full DFT to DFTB one loses quite some accuracy both on the PES, impacting on predicted absolute band positions, and on the dipole moment surface, resulting in much less accurate band intensities. Comparing in Fig.~\ref{fig:pyrene_anh_demon} the DFT-AC spectrum at the 523~K with the DFTB-MD one at the same temperature, we can see that the overall spectral pattern is similar but band positions given by DFTB-MD are clearly shifted with respect to DFT-AC, which is instead in very good agreement with experimental absorption spectra (see Table~\ref{Tab:anh_shift}). 
We can see that a bit less than half of the listed bands appear to lose intensity with increasing temperature, in the calculated spectra. This is due to the effect, that we described in Section~\ref{sec:results}, of some of the intensity of main bands being ``spread'' over an increasingly large number of very weak bands, forming a broad plateau below them which, at 523\,K, merges in a sort of featureless continuum. The weakest of these hot bands are not even recorded by the code, falling below the intensity threshold to save them. This artifact results in the total integrated intensity over the whole vibrational spectrum to decrease by almost 20\% between 14\,K and 523\,K.

Concerning the evolution of band positions and widths with temperature, we note first that we cannot give a ``rule of the thumb'' in the temperature interval explored by our DFT-AC calculations which would be valid for all bands in a given energy range: there are e.~g. some notable cases of bands that remain much sharper (more than an order of magnitude) than neighbouring ones. One such example is apparent in Figure~\ref{fig:ac_stick3}. Clearly, in these cases, none of the vibrational modes that can be populated at $\sim$523~K have an important anharmonic coupling with that band, at least up to 523~K. \\ 
A second observation is that the behaviour of band positions and widths (Figures~\ref{fig:bp_T_anh_demon_1} and \ref{fig:bw_T_anh_demon_2}) appears not linear at low temperatures in the spectra obtained via a fully quantum anharmonic calculation, in agreement with available laboratory data \citep{chakraborty2019}. This can be rationalised by considering that, at any given temperature T, vibrational states whose frequency is much larger than $k T / h$ are not populated, and thus do not participate at all in the structure of hot bands collectively producing band broadening and shift. If a vibrational mode has significant anharmonic coupling with a given band, it will produce a rather abrupt change in the evolution of the band when the temperature becomes sufficient to populate it. In a temperature range in which all modes with large anharmonic coupling with a band are already populated, and no new important modes come into play, band position and width instead evolve linearly with temperature. 
This accounts for the observed behaviour in Figures~\ref{fig:bp_T_anh_demon_1}, \ref{fig:bw_T_anh_demon_2}, and figures in the Supplementary Information, which appears to resemble broken lines. It also explains why, at high enough temperatures, all trends appear linear, as observed in experimental data \citep{Joblin1995, chakraborty2019}. \\
Classical molecular dynamics cannot account for some modes being completely depopulated at low temperature, since they do not consider quantised states and include vibrations of arbitrarily small amplitude, forbidden by quantum mechanics. As a result, these simulations \textit{always} predict a linear behaviour of band positions and widths, regardless of temperature. This behaviour is expected to approach the quantum one when vibrational energies are large enough. \\

\section{The combined view for astrophysical applications}\label{sec:combined}

Addressing the vibrational spectral evolution over a wide temperature range, extending up to $\sim$1500-2000\,K is a challenging task, but it is necessary to properly model PAH emission bands, including the effects of anharmonicity, in astronomical environments.  While we benchmark our theoretical methods against absorption spectra, we emphasize that our findings are perfectly applicable to the expected behaviour of band positions and spectral profiles in emission. In fact, emission spectra at a given temperature can be obtained from absorption spectra at the same temperature using Kirchoff's law. And since Planck's blackbody function is relatively slowly varying on the scale of the width of any PAH band, band position, width, and spectral profile can be considered to be identical, for the same band, with negligible error, between absorption and emission.

Our combined view using DFT-AC and DFTB-MD is restricted to quantify the \textit{evolution} with temperature of band positions and widths regardless of their absolute positions and intensities.
The data reported in Figures~\ref{fig:bp_T_anh_demon_1} and~\ref{fig:bw_T_anh_demon_2} (and Figures~S1, S2, S3, and S4 in Supplementary Information) show that both positions and widths follow different trends in different temperature ranges. This is in agreement with what was observed in laboratory experiments \citep{chakraborty2019}.
Band positions are fairly stable below $\sim$300\,K, and have an approximately linear asymptotic behaviour at high temperatures. As to bandwidths, they also have an approximately linear asymptotic behaviour at high temperatures, and some of them are relatively stable, like positions, below $\sim$300\,K, while others keep decreasing down to very low temperatures, but with a different slope. 
There are no ``one size fits all'' simple rules that can be used to predict anharmonic band positions and widths for all bands: the behaviour of individual bands must be investigated, by theoretical and/or laboratory experiments. Current models using a fixed shift with respect to 0\,K calculations, and/or fixed ``standard'' band widths to compute the emission of ``astronomical'' PAHs do not seem therefore appropriate. Their inaccuracy is masked, when comparing with astronomical observations, by the unresolved contribution of many different (unknown) individual molecules to astronomical AIBs, that does not allow one to separate the contribution of anharmonicity from those of chemical diversity and rotational broadening. On the other hand, the inaccuracy becomes apparent if one applies models to cases in which laboratory data of individual molecules are available, as we did here.\\
\citet{mackie2018}, proposed a simplified scheme for the calculation of simulated PAH emission spectra for comparison with AIBs, based on detailed theoretical modelling of cascade spectra that hinted that most bands did not show apparent shifts at different excitation energies, and suggested that in most cases only variations in bandwidths needed to be included for astronomical purposes. However, this conclusion appears difficult to reconcile with the linear shift of all band positions at high temperatures observed in laboratory spectra, and reproduced by theoretical modelling in the present work.

We therefore review our options, in view of their application to astronomical simulations, to assess the respective merits and drawbacks of the complementary theoreticals frameworks we presented.
When compared to laboratory data, we found that at frequencies higher than 1300\,cm$^{-1}$, the values of $\chi'$ from DFTB-MD appear to be systematically a factor $\sim$3 times, whereas they are well in line with experimental data for the bands below 1300\,cm$^{-1}$. This is consistent with the previous findings of \citet{Joalland2010} and \citet{Simon2011}. On the opposite, except for a couple of bands, DFT-AC somewhat underestimates the $\chi'$ values. \\
The absolute agreement of bandwidths computed by DFT-AC with corrected laboratory data points is surprising (cf. Figure~\ref{fig:bw_T_anh_demon_2}). In the case of band~11, the slope of the experimental points is markedly larger than the slope of DFT-AC points at lower temperature, hinting that it may change just around $\sim$550\,K. A closer examination of Figure~\ref{fig:ac_stick2}, indeed, shows several strong hot bands beginning to appear around the main band at around $\sim$500\,K. Their increased relative contribution to the band is thus likely to cause a jump in the bandwidth. The comparison in Figure~\ref{fig:bw_T_anh_demon_2} finally shows that DFTB-MD overestimates bandwidths at the highest frequencies.
Table~\ref{Tab:anh_fact_chi_dp} shows that the values of $\chi''$ deduced from DFTB-MD results, even if qualitatively consistent with experiment, appear less accurate than DFT-AC in the very few cases in which comparison is possible. More experimental data would be needed to draw firm conclusions on this.
The overall picture is consistent with DFTB-MD behaving worse with unresolved band clusters possibly involving resonances and better with isolated bands, which is the case for pyrene at low frequencies. A tentative explanation (see Section~\ref{sec:dftbmd_results}) is that when several unresolved transitions contribute to an unresolved band, the large \textit{intensity} errors due to DFTB reflects in a large error in the average position resulting from their blend.

Based on our modelling, and on available laboratory data, we therefore suggest the following relatively simplified recipe for fast, efficient, yet sufficiently accurate astronomical modelling of PAH emission, in the framework of the thermal approximation:
\begin{itemize}
    \item below 300\,K assume band positions to be approximately constant; interpolate linearly bandwidths between reference values at $\sim$300\,K and very low temperature (like our 14\,K data points); use as a reference the values (given in Tables~\ref{Tab:anh_fact} and \ref{Tab:anh_fact_chi_dp} for pyrene) obtained from either laboratory data or detailed simulation at these temperatures;
    \item above 300\,K assume band positions and widths evolve following a linear trend, whose slopes $\chi'$ and $\chi''$ can be obtained by the best available data: experimental, detailed modelling (such as e.~g. DFT-AC here or SPECTRO \citep{mackie2016}), molecular dynamics based on full DFT or better, molecular dynamics based on faster approximate methods (such as DFTB used here); this choice should be an educated guess on a case-by-case basis, since e.~g. DFTB-MD appears to provide fairly good results for strong, isolated bands, and conversely bad ones where band clusters merge;
    \item estimate the statistical thermal distribution of the given PAH as a result of the radiation field it is subjected to \citep[see e.~g. ][]{verstraete2001}, and obtain the resulting spectrum as an appropriate weighted average of spectra at each temperature estimated following the two above points.
\end{itemize}
We are in the process of developing such a model, in view of the forthcoming data from the James Webb Space Telescope.

\section{Conclusions}\label{sec:conclusions}

The difficulty to directly measure in laboratory experiments the vibrational emission of isolated PAHs excited by the absorption of UV photons, in order to study astronomical AIBs, has led to develop models to simulate the IR cooling cascades of PAHs including anharmonic effects \citep{Cook98_model,verstraete2001,Pech2002}. Such modelling must be as simple as possible, to be applicable in a systematic way, yet include enough details to achieve sufficient accuracy. These details, the inputs of the models, must be as thoroughly validated against experiments as possible, testing them against all available laboratory data. \\ 
Now, with the James Webb Space Telescope expected to be coming online in the near future, the concept of ``sufficient accuracy'' related to PAH emission models needs to be revisited, especially if we want to extract the maximum information from the unprecedented level of spectral and spatial details it will make available. This is the right time to extend models to really include anharmonicity in a \textit{systematic} way. \citet{mackie2018} showed that they can perform detailed calculations, using a method similar to our AnharmoniCaOs. However, as with AnharmoniCaOs, fully modelling anharmonic PAH emission is a computationally expensive undertaking for every single molecule, which does not make it easily applicable in a systematic way to existing PAH databases \citep{Malloci2007, bauschlicher2018}, in view of using them to interpret the detailed, subtle spectral structures that JWST data will unveil. On the other hand, cruder approximations, such as disregarding the dependence of band positions on excitation, even at the very high vibrational temperatures reached by an isolated PAH after the absorption of an UV photon, appear incompatible with already available laboratory data, as we have shown here. We thus propose here a recipe for an intermediate level of detail, which would be computationally less expensive than an all-out full-quantum anharmonic calculation, and thus more suitable for systematic use. This recipe consists on assuming that we can describe band positions and widths as broken lines, with a first linear section defined by a data point at very low temperature and one at $\sim$300\,K, and another linear trend, defined by asymptotic slopes $\chi'$ and $\chi''$, at higher temperatures, in agreement with all available experimental data. 
We showed that we can have several ways to access $\chi'$ and $\chi''$ for all bands of a given species, and we can combine them to incrementally improve modelling, going from the cheapest and less accurate DFTB-MD, to the more expensive fully quantum treatment of DFT-AC or similar codes \citep[e.g. Spectro][]{mackie2016}, to the most accurate, but limited in number of studied species, direct gas-phase laboratory measurement. \\
The natural development of this work will be to fill in anharmonic calculations (or suitable laboratory measurements), at least at $\sim$14\,K and $\sim$300\,K, of absolute band positions and widths, and of $\chi'$ and $\chi''$ values, in a systematic way, for species in available PAH spectral databases \citep{Malloci2007, bauschlicher2018}. We are already in the process of developing an anharmonic PAH emission model that makes use of this information to improve the accuracy of its predictions. In parallel, with the natural increase of available computational resources, we will extend DFT-AC calculations to higher temperatures, to check whether we already achieved the asymptotic regime at $\sim$523\,K and/or at what temperature the accuracy of the quartic expansion of the PES it employs begins to break down.\\
The ultimate test will be to compare the simulated spectra from anharmonic AIB models with direct measurements of UV-excited PAH infrared emission. These are very complex experiments and no progress could be achieved since the pioneering experiments of the 1990s \citep{Cherchneff1989,Shan1991,Brenner1992,Schlemmer94,Cook98}. While awaiting for further experimental results, the coming JWST data is a strong motivation on its own to proceed with this modelling effort.\\

\section*{Acknowledgements}

The research leading to these results was supported by the funding received from European Research Council under the European Union's seventh framework program (FP/2007-2013) ERC-2013-SyG, Grant agreement n. 610256 NANOCOSMOS. It was also supported by the CNES in the framework of the \textit{LAIBrary} project (ID 5830), in preparation of the JWST mission. \\
We acknowledge high performance computing resources made available by the “Accordo Quadro INAF-CINECA (2018)” and by CALMIP, Toulouse (project number: P16045). \\
We thank Cyril Falvo for support in obtaining the DFT data, as well as for providing the Wang-Landau sampling, that we used for our AnharmoniCaOs runs. \\
Finally, CJ acknowledges enthusiastic discussions with Stephan Schlemmer on UV-excited PAH experiments; these discussions have been alive since a common post-doctoral time at the bay area (CA).

\section*{Dataset}
The dataset associated with this work can be found on the Zenodo platform~\citep{zenodo_2021} and via the soon to be released cosmic-pah portal at \mbox{https://cosmic-pah.irap.omp.eu}.

\bibliography{mybibfile}

\end{document}


\begin{frontmatter}

\title{Supporting information for:\\Anharmonic Infrared Spectra of Thermally Excited Pyrene (C$_{16}$H$_{10}$): A Combined View of DFT-Based GVPT2 with AnharmonicCaOs, and Approximate DFT Molecular dynamics with DemonNano}


\author[mymainaddress]{Shubhadip Chakraborty}

\author[mymainaddress,mysecondaryaddress]{Giacomo Mulas\corref{mycorrespondingauthor}}

\author[mysecondaryaddress1]{Mathias Rapacioli}

\author[mymainaddress]{Christine Joblin}

\cortext[mycorrespondingauthor]{Corresponding author}

\address[mymainaddress]{Institut de Recherche en Astrophysique et Plan\'etologie, Universit\'e de Toulouse (UPS), CNRS, CNES, 9 Av. du Colonel Roche, 31028 Toulouse Cedex 4, France.}
\address[mysecondaryaddress]{Istituto Nazionale di Astrofisica (INAF), Osservatorio Astronomico di Cagliari, 09047 Selargius (CA), Italy}
\address[mysecondaryaddress1]{Laboratoire de Chimie et Physique Quantiques (LCPQ/IRSAMC), Universit\'e de Toulouse (UPS),CNRS, 118 Route de Narbonne, 31062 Toulouse, France}

\end{frontmatter}

\linenumbers
\begin{table}[H]
\centering
\footnotesize
\caption{\textbf{Integration range for deriving band positions in spectra generated by AnharmoniCaOs package}}
\hspace{2cm}
\begin{tabular}{c c c}
\hline
\multicolumn{1}{c}{No}&\multicolumn{2}{c}{AnharmoniCaOs}\\
\cline{2-3}
\multicolumn{1}{c}{}&\multicolumn{1}{c}{Position at 300\,K}&\multicolumn{1}{c}{Range}\\
\hline
1 & 3040 &3060-3020 \\
2 & 1585 &1605-1656 \\
3 & 1473 & 1493-1453 \\
4 & 1424 & 1444-1404\\
5 & 1311 & 1331-1291 \\
6 & 1241 & 1261-1221 \\
7 & 1180 & 1200-1160 \\
8 & 1089 & 1109-1069 \\
9 & 996 & 1016-976 \\
10 & 964 & 984-944 \\
11 & 843 & 863-823 \\
12 & 816 & 836-796 \\
13 & 739 & 760-730 \\
14 & 710 & 730-690  \\
\hline
\label{Tab:range_AC}
\end{tabular}
\end{table}

\begin{table}[H]
\centering
\footnotesize
\caption{\textbf{Integration range for deriving band positions in spectra generated by deMonNano package}}
\hspace{2cm}
\resizebox{\textwidth}{!}{\begin{tabular}{c c c c c c c c c c}
\hline
\multicolumn{1}{c}{Band position}&\multicolumn{8}{c}{Temperature (K)}\\
\cline{2-10}
\multicolumn{1}{c}{at 300\,K}&\multicolumn{1}{c}{200}&\multicolumn{1}{c}{300}&\multicolumn{1}{c}{400}&\multicolumn{1}{c}{600}&\multicolumn{1}{c}{800}&\multicolumn{1}{c}{1000}&\multicolumn{1}{c}{1200}&\multicolumn{1}{c}{1400}&\multicolumn{1}{c}{1600}\\
\hline
723.7&740 - 715&740 - 715&740 - 712&740 - 709&740 - 706&740 - 703&740 - 700&740 - 697&740 - 696\\
814.1& 830 - 770 & 830 -770 & 830 - 770 & 830 - 770 & 830 - 770 & 830 -770 & 830 - 770 & 830 -770 & 830 -770\\
1089.2 & 1098 - 1084 & 1095 - 1081 & 1092 - 1078 & 1090 - 1073 & 1088 - 1068 & 1084 - 1062 & 1082 - 1058 & 1080 - 1053 & 1078 - 1051 \\
1333.1 & 1354 - 1323 & 1354 - 1321 & 1354 - 1319 & 1354 - 1312 & 1354 - 1308 & 1354 - 1304 & 1354 - 1298 & 1354 - 1291 & 1354 - 1291 \\
1465.8 & 1481 - 1462 & 1476 - 1454 & 1468 - 1453 & 1458 - 1436 & 1453 - 1430 & 1446 - 1425 & 1440 - 1414 & 1430 - 1400 & 1430 - 1400 \\
1628.5 & 1643 - 1624 & 1640 - 1612 & 1637 - 1604 & 1631 - 1597 & 1623 - 1583 & 1617 - 1567 & 1603 - 1552 & 1591 - 1552 & 1580 - 1540 \\
1585.5 & 1597 - 1582 & 1592 - 1576 & 1590 - 1572 & 1583 - 1566 & 1568 - 1533 & 1566 - 1522 & 1561 - 1517 & 1547 -1506 & 1539 - 1500\\
1842.4 & 1862 - 1840 & 1858 - 1829 & 1855 - 1826 & 1847 - 1779 & 1835 - 1779 & 1829 - 1769 & 1813 - 1740 & 1804 - 1736 & 1804 - 1714\\
2923.9 & 3000 -2700 & 3000 - 2700 & 3000 - 2700 & 3000 - 2700 & 3000 - 2700 & 3000 - 2700 & 3000 - 2700 & 3000 - 2700 & 3000 - 2700\\
\hline
\label{Tab:range_MD}
\end{tabular}}
\end{table}

\begin{longtable}{c c c c c c}
\caption{\textbf{List of harmonic frequencies of pyrene using DFT, DFTB and anharmonic frequencies calculated using AnharmoniCaOs package at 0 K.}} \label{tab:py_freq} \\
\hline \multicolumn{1}{c}{Sym}&\multicolumn{1}{c}{No}&\multicolumn{2}{c}{DFT}&\multicolumn{2}{c}{DFTB}\\
& & Freq & Int & Freq & Int \\
& & (cm$^{-1}$) & (km mol$^{-1}$) & (cm$^{-1}$) & (km mol$^{-1}$) \\
\hline
\endfirsthead
\multicolumn{3}{c}%
{{\bfseries \tablename\ \thetable{} -- continued from previous page}} \\
\hline \multicolumn{1}{c}{Sym}&\multicolumn{1}{c}{No}&\multicolumn{2}{c}{DFT}&\multicolumn{2}{c}{DFTB}\\
& & Freq & Int & Freq & Int \\
& & (cm$^{-1}$) & (km mol$^{-1}$) & (cm$^{-1}$) & (km mol$^{-1}$) \\\hline
\endhead

\hline \multicolumn{4}{r}{{Continued on next page}} \\ \hline
\endfoot
\hline \hline
\endlastfoot
A$_{g}$&$\nu_{1}$ & 3182.2 & 0.000 & 2965.8 &	0.000 \\
B$_{1u}$&$\nu_{2}$ &	3182.0	&	37.998	&	2965.7	&	42.585\\
A$_{g}$&$\nu_{3}$ 	&	3172.8	&	0.000	&	2960.4	&	0.000\\
B$_{2u}$&$\nu_{4}$ 	&	3172.7	&	42.847	&	2960.3	&	56.295\\
B$_{3g}$&$\nu_{5}$ 	&	3165.5	&	0.000	&	2958.0	&	0.000\\
B$_{2u}$&$\nu_{6}$ 	&	3165.0	&	12.674	&	2958.0	&	36.845\\
B$_{1u}$&$\nu_{7}$ 	&	3157.7	&	0.002	&	2953.7	&	1.515\\
A$_{g}$&$\nu_{8}$ 	&	3157.5	&	0.000	&	2953.8	&0.000\\
B$_{3g}$&$\nu_{9}$ 	&	3154.6	&	0.000	&	2949.6	&	0.000\\
B$_{1u}$&$\nu_{10}$	&	3154.4	&	1.229	&	2949.7	&	7.515\\
A$_{g}$&$\nu_{11}$	&	1661.0	&	0.000	&	1863.7	&	0.000	\\
B$_{2u}$&$\nu_{12}$	&	1635.8	&	2.257	&	1856.3	&	9.175	\\
B$_{1u}$&$\nu_{13}$	&	1626.5	&	13.181	&	1836.5	&	2.487	\\
B$_{3g}$&$\nu_{14}$	&	1618.3	&	0.000	&	1814.9	&	0.000\\
A$_{g}$&$\nu_{15}$	&	1587.8	&	0.000	&	1797.1	&	0.000\\
B$_{3g}$&$\nu_{16}$	&	1528.9	&	0.000	&	1710.2	&	0.000\\
B$_{2u}$&$\nu_{17}$	&	1508.5	&	3.199	&	1699.9	&	0.001\\
B$_{1u}$&$\nu_{18}$	&	1478.9	&	0.993	&	1644.5	&	6.832\\
B$_{1u}$&$\nu_{19}$	&	1456.5	&	8.010	&	1576.1	&	0.588\\
B$_{2u}$&$\nu_{20}$	&	1453.4	&	3.197	&	1598.4	&	4.313\\
B$_{3g}$&$\nu_{21}$	&	1433.0	&	0.000	&	1506.9	&	0.000\\
A$_{g}$&$\nu_{22}$	&	1420.8	&	0.000	&	1601.9	&	0.000\\
B$_{3g}$&$\nu_{23}$	&	1393.9	&	0.000	&	1561.4	&	0.000	\\
A$_{g}$&$\nu_{24}$	&	1347.3	&	0.000	&	1465.7	&	0.000\\
B$_{2u}$&$\nu_{25}$	&	1338.3	&	4.911	&	1477.1	&	2.971	\\
B$_{1u}$&$\nu_{26}$	&	1264.3	&	2.261	&	1335.9	&	25.189	\\
B$_{3g}$&$\nu_{27}$	&	1260.7	&	0.000	&	1319.1	&	0.000	\\
A$_{g}$&$\nu_{28}$	&	1256.9	&	0.000	&	1390.9	&	0.000	\\
B$_{2u}$&$\nu_{29}$	&	1226.2	&	0.020	&	1318.9	&	9.862	\\
B$_{2u}$&$\nu_{30}$	&	1202.2	&	14.025	&	1236.3	&	0.235\\
B$_{3g}$&$\nu_{31}$	&	1194.9	&	0.000	&	1241.6	&	0.000\\
A$_{g}$&$\nu_{32}$	&	1164.9	&	0.000	&	1203.4	&	0.000\\
B$_{2u}$&$\nu_{33}$	&	1160.4	&	0.276	&	1200.3	&	1.192\\
B$_{3g}$&$\nu_{34}$	&	1122.9	&	0.000	&	1178.2	&	0.000	\\
B$_{1u}$&$\nu_{35}$	&	1109.2	&	7.048	&	1177.7	&	3.461	\\
A$_{g}$&$\nu_{36}$	&	1085.6	&	0.000	&	1169.1	&	0.000	\\
B$_{1u}$&$\nu_{37}$	&	1011.5	&	1.791	&	1094.2	&	1.120	\\
B$_{2g}$&$\nu_{38}$	&	990.7	&	0.000	&	967.3	&	0.000\\
A$_{u}$&$\nu_{39}$	&	985.1	&	0.000	&	962.3	&	0.000	\\
B$_{3u}$&$\nu_{40}$	&	983.8	&	1.367	&	953.8	&	0.035	\\
B$_{2g}$&$\nu_{41}$	&	977.0	&	0.000	&	953.2	&	0.000	\\
B$_{2u}$&$\nu_{42}$	&	975.4	&	0.003	&	1077.6	&	2.143	\\
B$_{1g}$&$\nu_{43}$	&	920.6	&	0.000	&	886.3	&	0.000	\\
A$_{u}$&$\nu_{44}$	&	908.2	&	0.000	&	879.3	&	0.000 \\
B$_{3u}$&$\nu_{45}$	&	860.0	&	111.896&	818.4	&	12.359	\\
B$_{2g}$&$\nu_{46}$	&	859.2	&	0.000	&	809.5	&	0.000	\\
B$_{1u}$&$\nu_{47}$	&	830.3	&	4.584	&	868.0	&	0.156\\
B$_{1g}$&$\nu_{48}$	&	815.4	&	0.000	&	788.3	&	0.000	\\
A$_{g}$&$\nu_{49}$	&	811.3	&	0.000	&	839.3	&	0.000	\\
B$_{2g}$&$\nu_{50}$	&	783.6	&	0.000	&	762.3	&	0.000	\\
B$_{3u}$&$\nu_{51}$	&	755.1	&	17.533&	725.1	&	1.641	\\
B$_{3g}$&$\nu_{52}$	&	745.2	&	0.000	&	778.0	&	0.000	\\
B$_{3u}$&$\nu_{53}$	&	725.6	&	44.795&	702.1	&	0.291	\\
B$_{1u}$&$\nu_{54}$	&	700.8	&	0.029	&	725.8	&	0.620	\\
A$_{u}$&$\nu_{55}$	&	689.7	&	0.000	&	669.9	&	0.000	\\
A$_{g}$&$\nu_{56}$	&	593.2	&	0.000	&	670.9	&	0.000	\\
B$_{2g}$&$\nu_{57}$	&	586.5	&	0.000	&	566.7	&	0.000	\\
B$_{2u}$&$\nu_{58}$	&	549.2	&	2.552	&	563.6	&	0.830	\\
B$_{1g}$&$\nu_{59}$	&	533.9	&	0.000	&	511.3	&	0.000	\\
B$_{2g}$&$\nu_{60}$	&	511.7	&	0.000	&	490.5	&	0.000	\\
B$_{3g}$&$\nu_{61}$	&	504.2	&	0.000	&	513.7	&	0.000	\\
B$_{1u}$&$\nu_{62}$	&	503.7	&	2.984	&	524.6	&	0.918	\\
B$_{3u}$&$\nu_{63}$	&	496.3	&	2.118	&	474.9	&	0.147	\\
B$_{3g}$&$\nu_{64}$	&	459.7	&	0.000	&	471.1	&	0.000	\\
A$_{g}$&$\nu_{65}$	&	409.6	&	0.000	&	418.4	&	0.000	\\
A$_{u}$&$\nu_{66}$	&	399.9	&	0.000	&	383.1	&	0.000	\\
B$_{2u}$&$\nu_{67}$	&	356.5	&	1.678	&	362.3	&	0.141	\\
B$_{2g}$&$\nu_{68}$	&	259.7	&	0.000	&	253.8	&	0.000	\\
B$_{1g}$&$\nu_{69}$	&	246.9	&	0.000	&	240.2	&	0.000	\\
B$_{3u}$&$\nu_{70}$	&	210.2	&	10.595&	208.7	&	0.075	\\
A$_{u}$&$\nu_{71}$	&	149.9	&	0.0	&	149.7	&	0.000	\\
B$_{3u}$&$\nu_{72}$	&	97.2	&	0.635	&	97.6	&	0.002	\\
\hline 
\end{longtable}

\newpage
\section{Illustration of the variation of band positions with temperature obtained from AnharmoniCaOs (DFT) and deMonNano (MD) simulation}

\textbf{AnharmoniCaOs}

\begin{figure}[H]
    \centering
    \includegraphics[width=1.0\columnwidth]{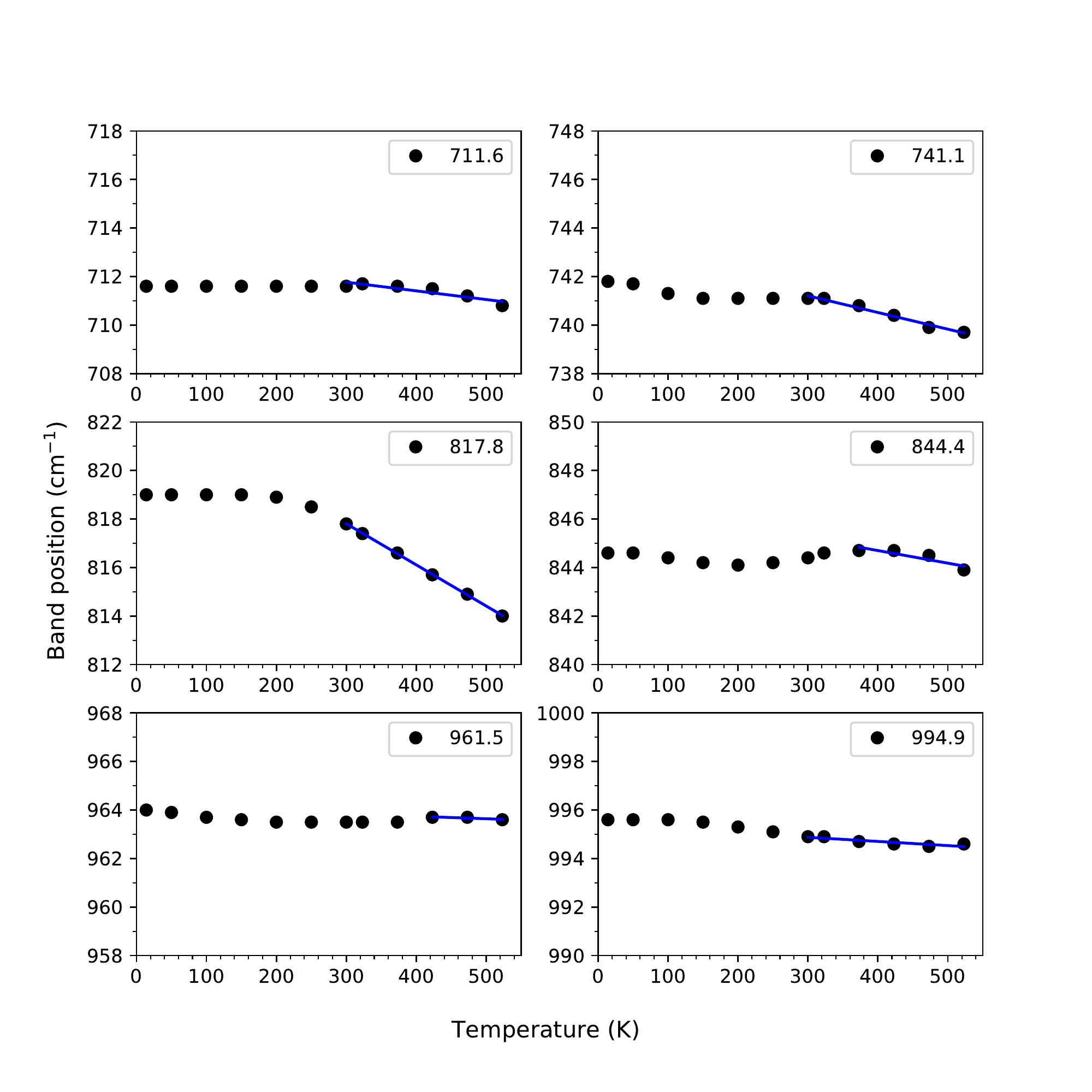}
    \caption{Evolution of band positions with temperature obtained from AnharmoniCaOs. The linear trend of the band position with temperature is applicable only for the high temperature range. Therefore only the points from and beyond 300\,K have been considered for the fitting. Band position at 300\,K are marked in each panel of the figure.}
    \label{fig:pyrene_bp1}
\end{figure}

\begin{figure}[H]
\ContinuedFloat
    \centering
    \includegraphics[width=1.0\columnwidth]{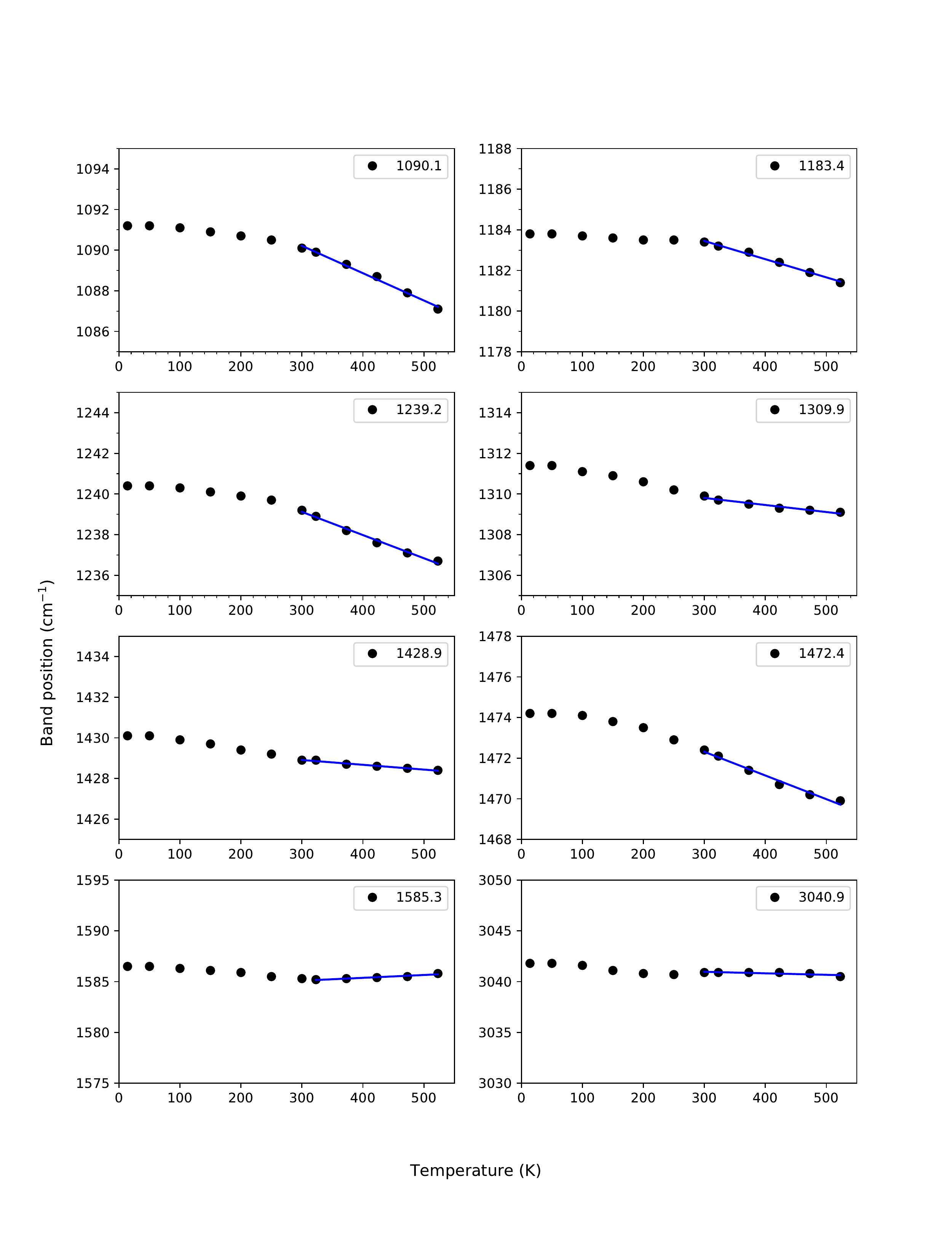}
    \caption{Evolution of band positions with temperature obtained from AnharmoniCaOs. The linear trend of the band position with temperature is applicable only for the high temperature range. Therefore only the points from and beyond 300\,K have been considered for the fitting. Band position at 300\,K are marked in each panel of the figure. (Continued from the previous page)}
    \label{fig:pyrene_bp2}
\end{figure}

\newpage
\textbf{deMonNano}

\begin{figure}[H]

    \centering
    \includegraphics[width=1.0\columnwidth]{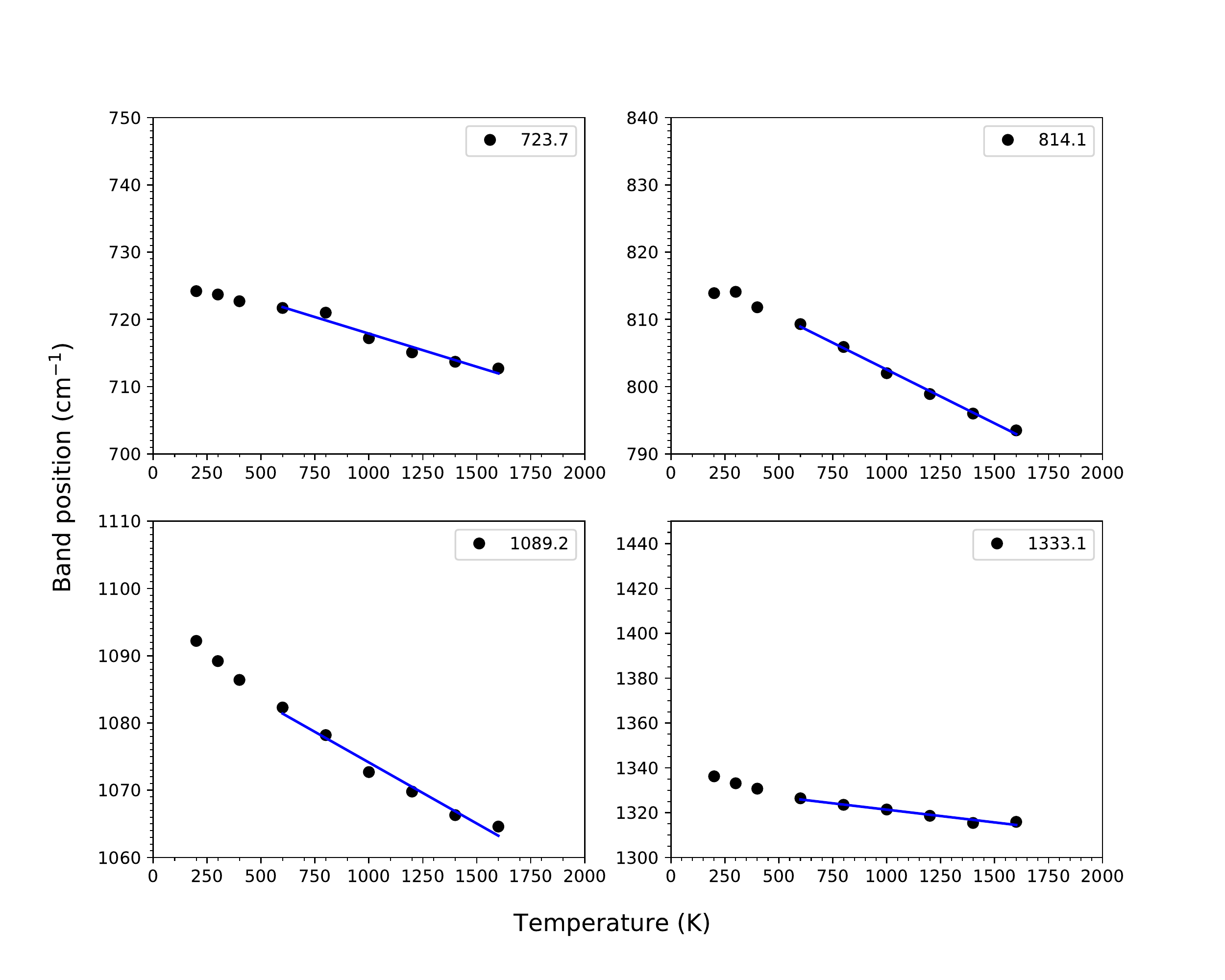}
    \caption{Evolution of band positions with temperature obtained from MD (demonNano) simulation. This method is dedicated to simulate the band positions at high temperature. Therefore only the points from and beyond 600\,K have been considered for the fitting. Band position at 300\,K are marked in each panel of the figure.}
    \label{fig:pyrene_bp_demon1}
\end{figure}

\begin{figure}[H]
\ContinuedFloat
    \centering
    \includegraphics[width=1.0\columnwidth]{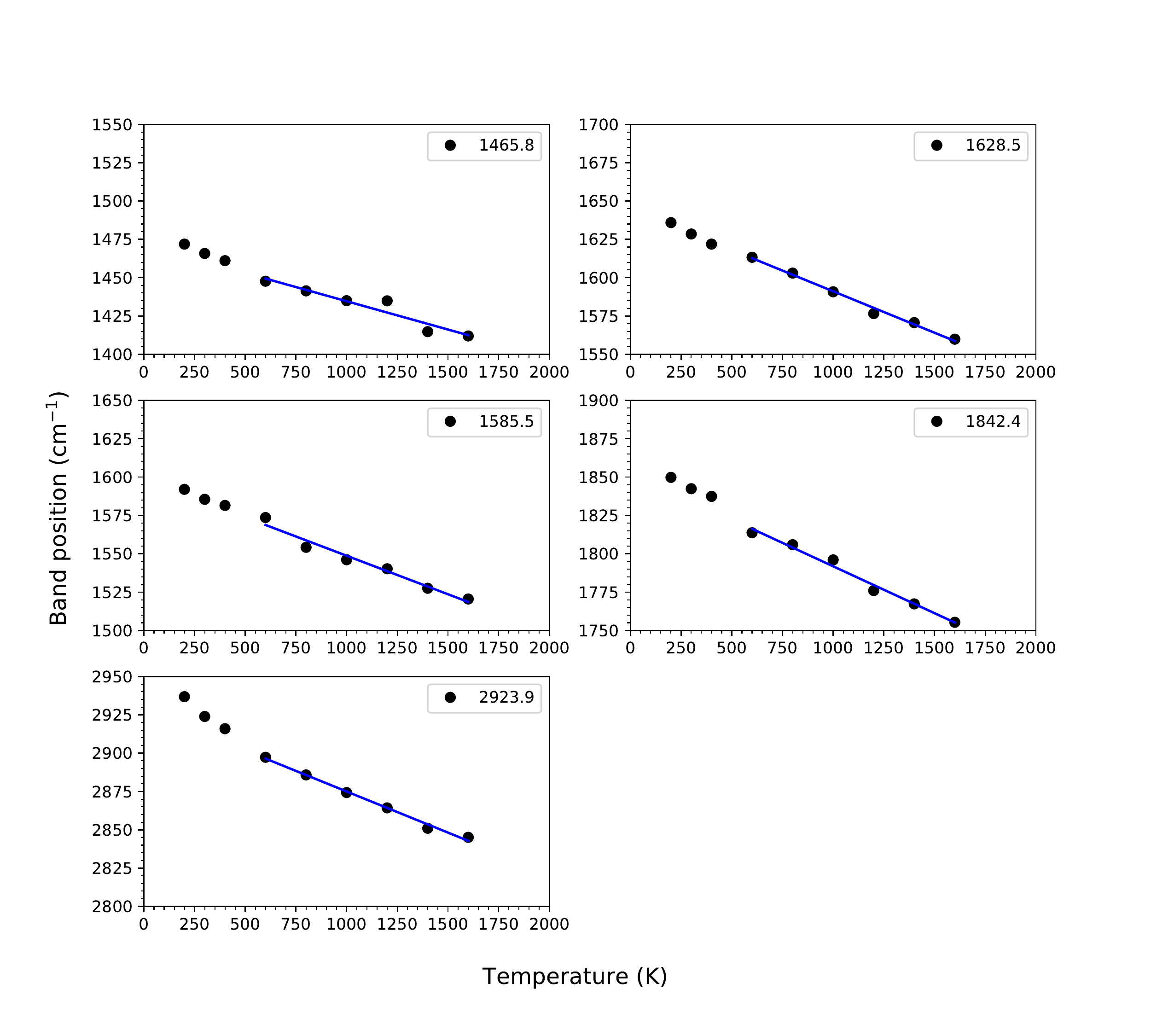}
    \caption{Evolution of band positions with temperature obtained from MD (demonNano) simulation. This method is dedicated to simulate the band positions at high temperature. Therefore only the points from and beyond 600\,K have been considered for the fitting. Band position at 300\,K are marked in each panel of the figure. (Continued from the previous page)}
    \label{fig:pyrene_bp_demon2}
\end{figure}

\section{Illustration of the variation of bandwidths with temperature obtained from AnharmoniCaOs (DFT) and deMonNano (MD) simulation}

\textbf{AnharmoniCaOs}

\begin{figure}[H]
    \centering
    \includegraphics[width=1.0\columnwidth]{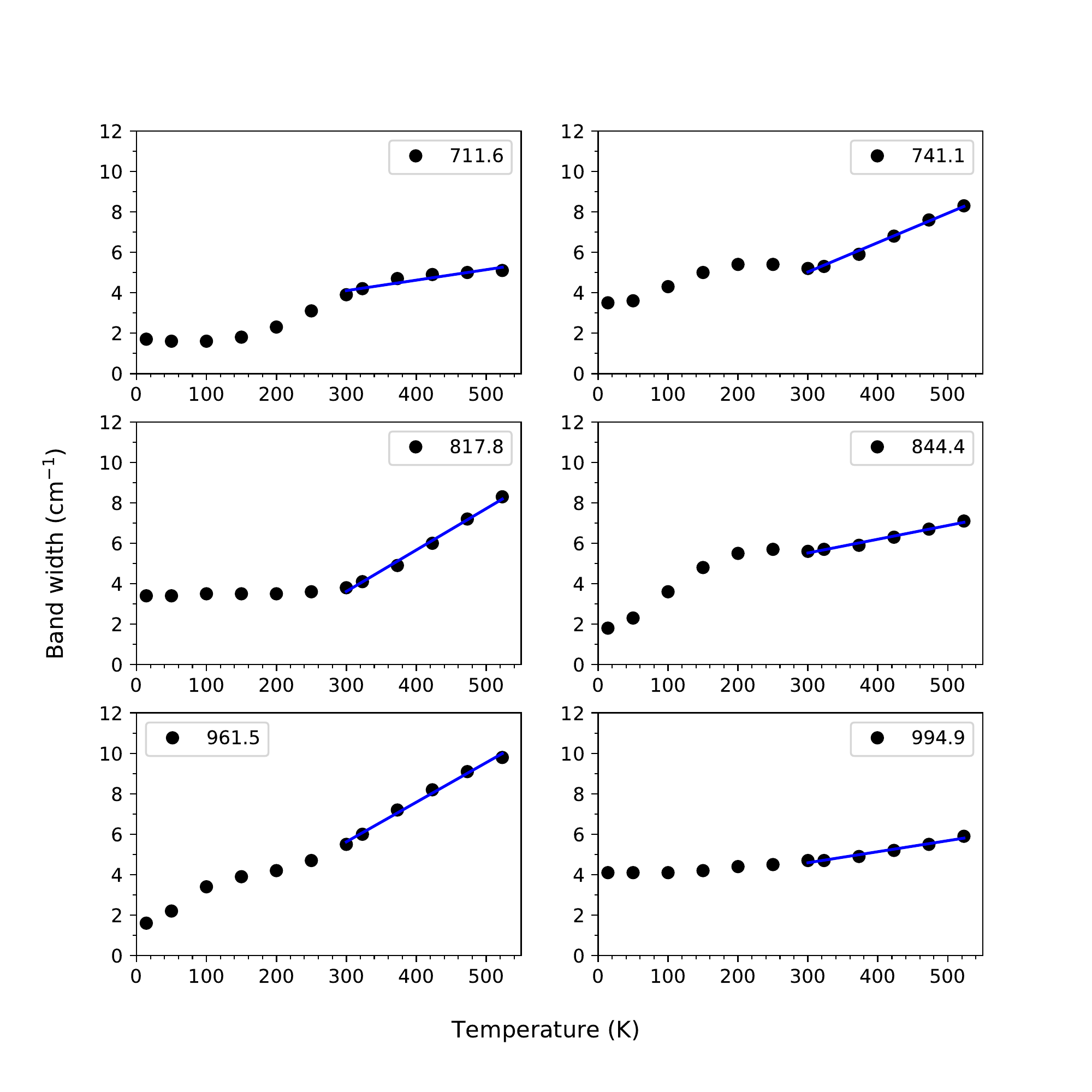}
    \caption{Evolution of bandwidths with temperature obtained from AnharmoniCaOs. The linear trend of the band position with temperature is applicable only for the high temperature range. Therefore only the points from and beyond 300\,K have been considered for the fitting. Band position at 300\,K are marked in each panel of the figure.}
    \label{fig:pyrene_bw1}
\end{figure}

\begin{figure}[H]
\ContinuedFloat
    \centering
    \includegraphics[width=1.0\columnwidth]{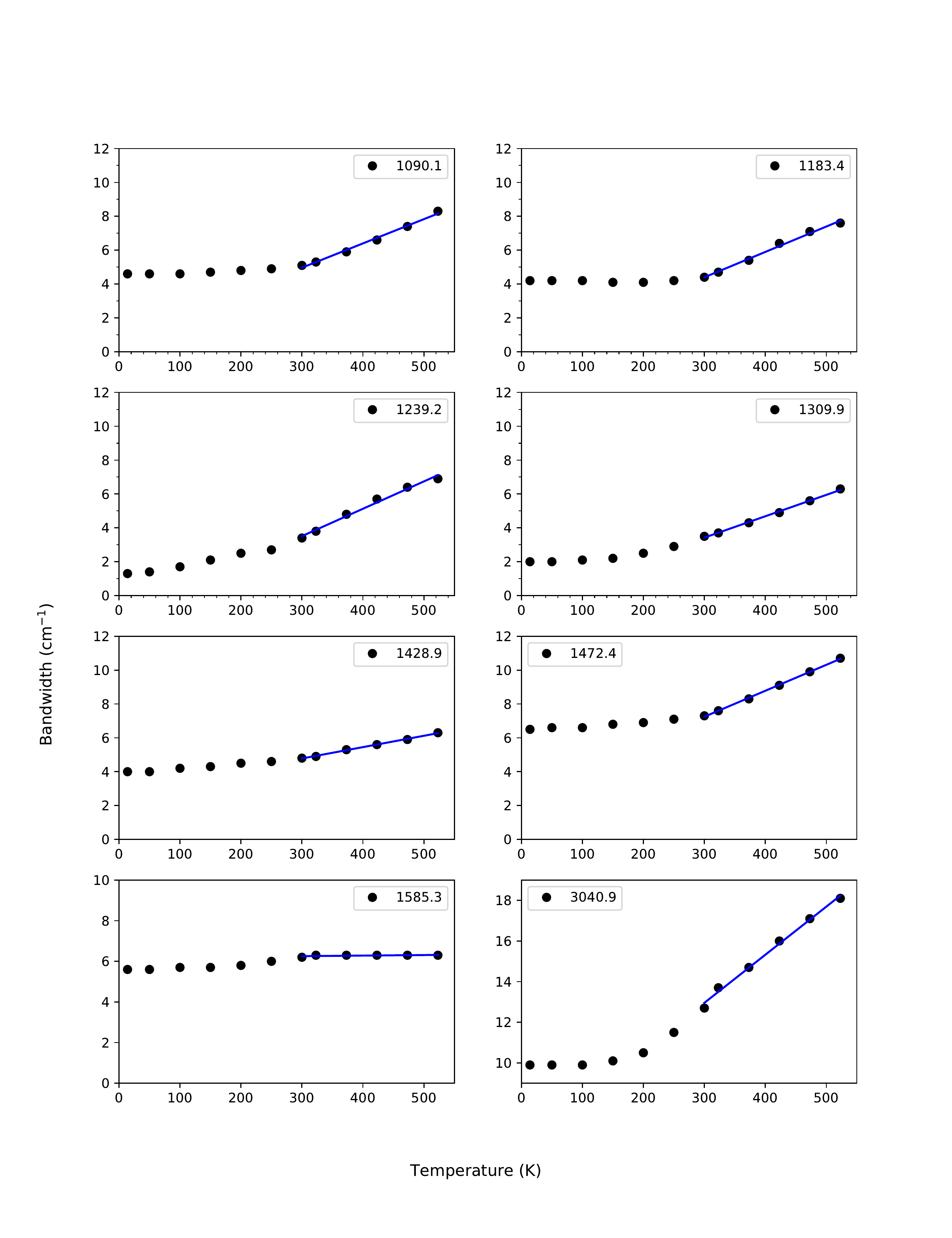}
    \caption{Evolution of bandwidths with temperature obtained from AnharmoniCaOs. The linear trend of the band position with temperature is applicable only for the high temperature range. Therefore only the points from and beyond 300\,K have been considered for the fitting. Band position at 300\,K are marked in each panel of the figure. (Continued from the previous page)}
    \label{fig:pyrene_bw2}
\end{figure}

\newpage
\textbf{deMonNano}

\begin{figure}[H]
    \centering
    \includegraphics[width=1.0\columnwidth]{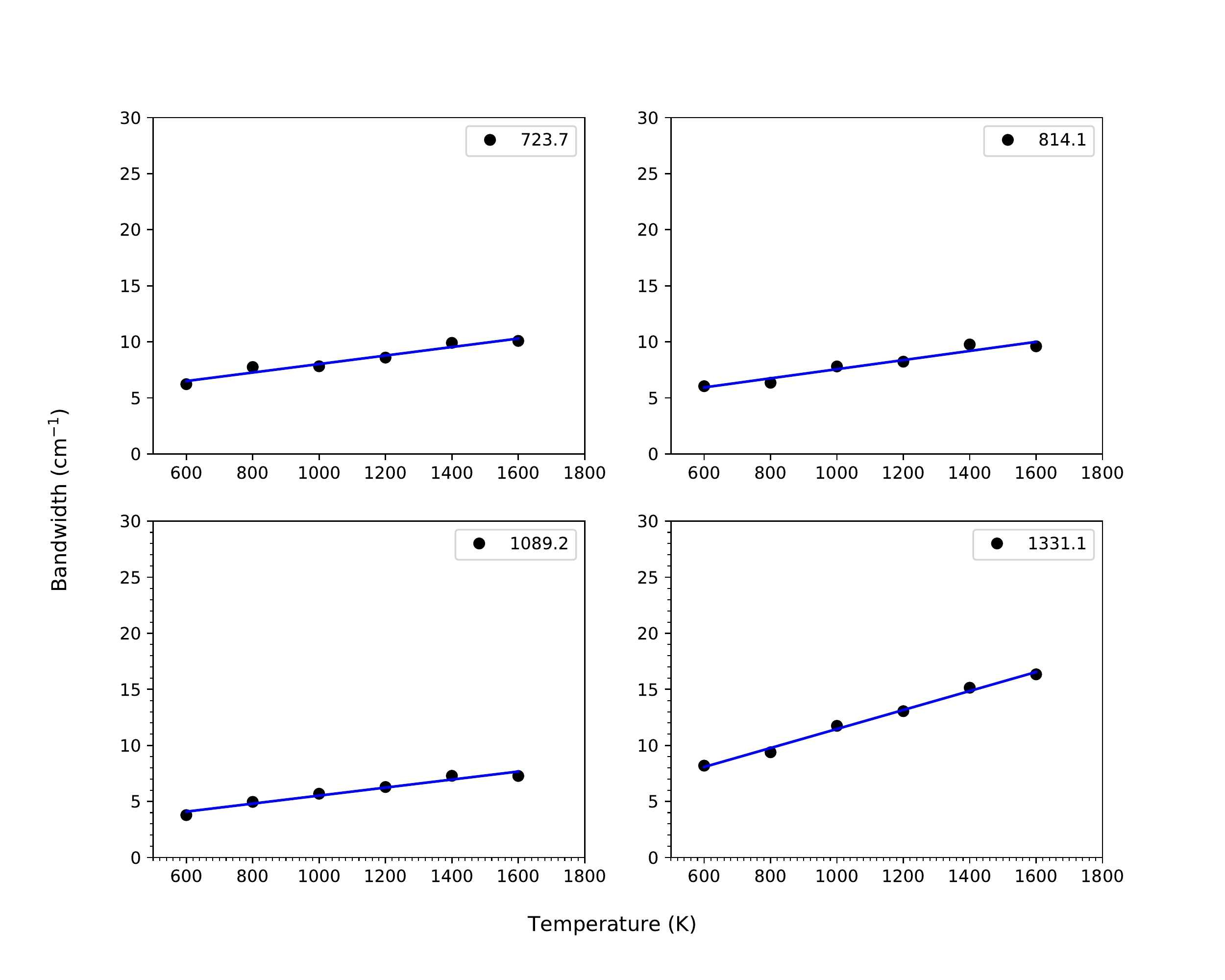}
    \caption{Evolution of bandwidths with temperature obtained from MD (demonNano) simulation. This method is dedicated to simulate the band positions at high temperature. Therefore only the points from and beyond 600\,K have been considered for the fitting. Band position at 300\,K are marked in each panel of the figure.}
    \label{fig:pyrene_bw1_demon}
\end{figure}

\begin{figure}[H]
\ContinuedFloat
    \centering
    \includegraphics[width=1.0\columnwidth]{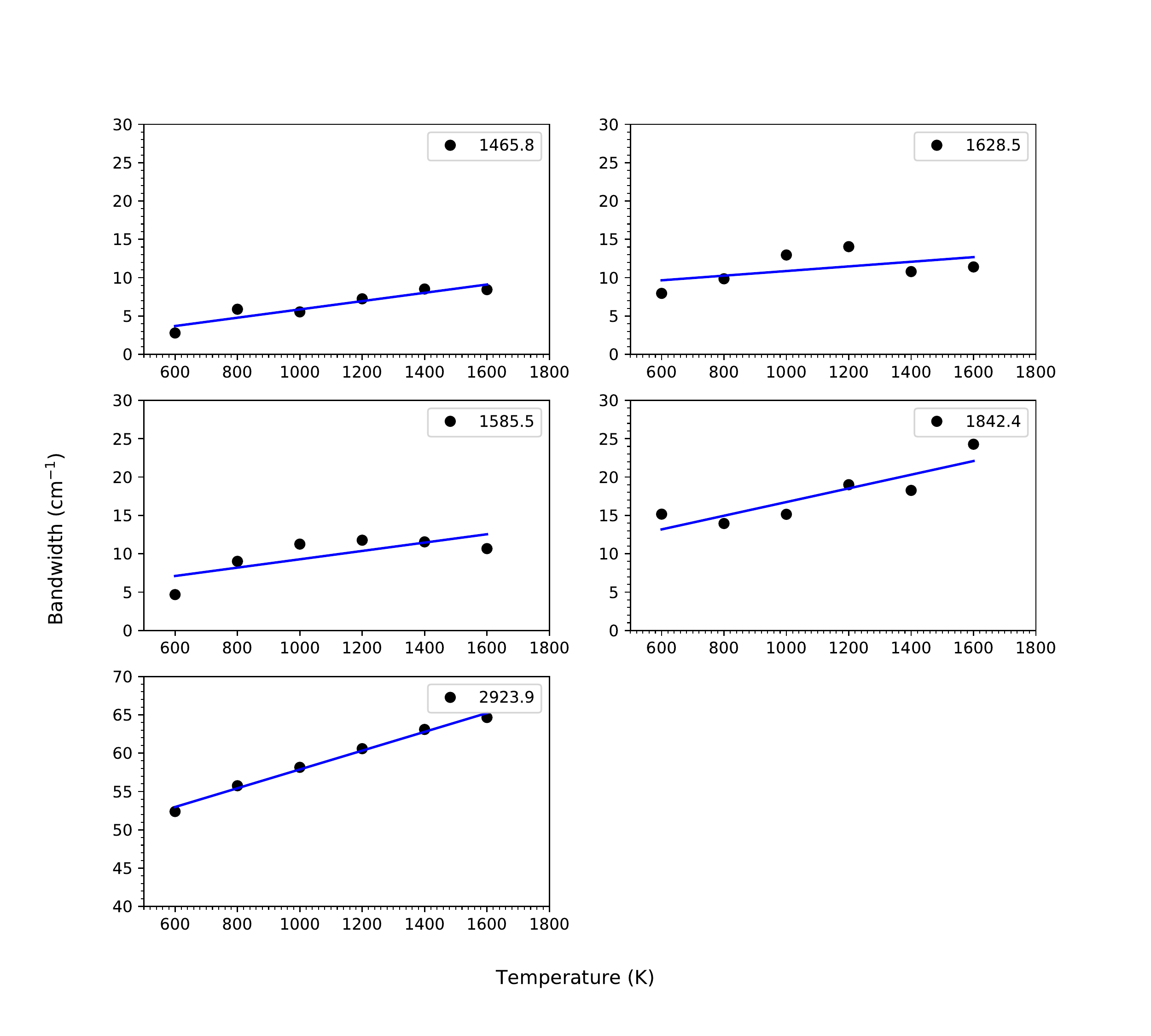}
    \caption{Evolution of bandwidths with temperature obtained from MD (demonNano) simulation. This method is dedicated to simulate the band positions at high temperature. Therefore only the points from and beyond 600\,K have been considered for the fitting. Band position at 300\,K are marked in each panel of the figure. (Continued from the previous page)}
    \label{fig:pyrene_bw2_demon}
\end{figure}